\documentclass[prd,aps,preprintnumbers,nofootinbib,superscriptaddress]{revtex4}

\usepackage{amssymb,amsmath,graphicx,bm}

\allowdisplaybreaks
\usepackage[usenames,dvipsnames]{color}
\usepackage[normalem]{ulem} 
\usepackage{slashed} 
\renewcommand\sout{\bgroup \color[rgb]{0.55,0.00,0.99} \ULdepth=-.5ex \ULset}
\newcommand{\xB }{x_{\scriptscriptstyle B}}
\newcommand{\sT}{{\scriptscriptstyle T}}
\renewcommand{\d}{\mathrm{d}}
\def\slash#1{\setbox0=\hbox{$#1$}               
        \dimen0=\wd0                            
        \setbox1=\hbox{/} \dimen1=\wd1          
        \ifdim\dimen0>\dimen1                   
        \rlap{\hbox to \dimen0{\hfil/\hfil}}    
        #1                                      
        \else
        \rlap{\hbox to \dimen1{\hfil$#1$\hfil}} 
        /                                       
        \fi}                                    %

\begin{document}

\title{Azimuthal asymmetries in semi-inclusive $J/\psi\,+\,\mathrm{jet}$ production at an EIC}

\author{Umberto D'Alesio}
\email{umberto.dalesio@ca.infn.it}
\affiliation{Dipartimento di Fisica, Universit\`a di Cagliari, Cittadella Universitaria, I-09042 Monserrato, Cagliari, Italy}
\affiliation{INFN, Sezione di Cagliari, Cittadella Universitaria, I-09042 Monserrato, Cagliari, Italy}

\author{Francesco Murgia}
\email{francesco.murgia@ca.infn.it}
\affiliation{INFN, Sezione di Cagliari, Cittadella Universitaria, I-09042 Monserrato, Cagliari, Italy}

\author{Cristian Pisano}
\email{cristian.pisano@ca.infn.it}
\affiliation{Dipartimento di Fisica, Universit\`a di Cagliari, Cittadella Universitaria, I-09042 Monserrato, Cagliari, Italy}
\affiliation{INFN, Sezione di Cagliari, Cittadella Universitaria, I-09042 Monserrato, Cagliari, Italy}

\author{Pieter Taels}
\email{pieter.taels@ca.infn.it}
\affiliation{INFN, Sezione di Cagliari, Cittadella Universitaria, I-09042 Monserrato, Cagliari, Italy}

\begin{abstract}
\end{abstract}
\date{\today}

\begin{abstract}
We consider transverse momentum dependent gluon distributions inside both unpolarized and transversely polarized protons and show how they can be probed by looking at azimuthal modulations in $e \, p \to  e \, J/\psi \,  \mathrm{jet} \, X$. We find that the contribution due to quark induced subprocesses is always suppressed in the considered kinematic regions, accessible in principle at a future Electron-Ion Collider. Our model-independent estimates of the maximal values of these asymmetries allowed by positivity bounds suggest the feasibility of their measurement. In addition, by adopting the McLerran-Venugopalan model for the unpolarized and linearly polarized gluon densities, we study the behavior of the
$\cos2\phi$ asymmetries in the small-$x$ limit.

\end{abstract}

\maketitle

\section{Introduction}
Transverse momentum dependent distribution functions (TMDs) of gluons inside unpolarized and polarized protons have been defined  in terms of QCD operators for the first time in Refs.~\cite{Mulders:2000sh,Meissner:2007rx}. Since then, they have received growing attention, mainly because they encode essential information on the transverse motion of gluons and their spin-orbit correlations. As such, they parameterize highly nontrivial features of the partonic structure of the proton. For instance, the distribution of linearly polarized gluons can be nonzero even if the parent proton is unpolarized and, if sizable, can modify the transverse spectra of (pseudo)scalar particles like, for example, the Higgs boson~\cite{Sun:2011iw,Boer:2011kf,Boer:2013fca,Echevarria:2015uaa,Gutierrez-Reyes:2019rug}. Another example is provided by the gluon Sivers function~\cite{Sivers:1989cc,Boer:2015ika}. In general, the Sivers function describes the azimuthal distributions of unpolarized partons inside a proton that is transversely polarized with respect to its momentum; it is expected to generate observable single spin asymmetries in processes initiated by transversely polarized protons. Moreover, it can provide an indication on how much quarks and gluons contribute to the total spin of the proton through their orbital angular momentum.

Gluon TMDs are also of great  interest because of their intrinsic process dependence, due to  their gauge link structure. This is determined by soft gluon exchanges occurring in the particular processes in which the TMDs are probed.  Unambiguous tests of such properties have been proposed for both the quark~\cite{Brodsky:2002rv,Collins:2002kn}  and the gluon~\cite{Boer:2016fqd} Sivers distributions.  Their verification will represent an important confirmation of our present knowledge of the TMD formalism and nonperturbative QCD in general.

Experimentally, not much is known about gluon TMDs. However, many proposals have been suggested to probe them, especially through the measurement of transverse momentum distributions and azimuthal asymmetries for heavy-quarkonium production, both in lepton-proton~\cite{Yuan:2008vn,Mukherjee:2016qxa,Rajesh:2018qks,Bacchetta:2018ivt} and in proton-proton collisions~\cite{Boer:2012bt,Dunnen:2014eta,Lansberg:2017tlc,Lansberg:2017dzg,DAlesio:2017rzj,DAlesio:2018rnv}.
In particular, in a very recent publication~\cite{Bacchetta:2018ivt}, some of us considered the inclusive production of a $J/\psi$ or a $\Upsilon$ meson in deep-inelastic electron-proton scattering, namely $e\,p\to e\,J/\psi\,(\Upsilon)\,X$. The corresponding cross section can be calculated in the TMD formalism only when two well-separated scales are present: a soft one, sensitive to the intrinsic parton transverse momenta, and a hard one, which allows for a perturbative treatment. Hence the analyzed kinematic region was the one in which the transverse momentum $q_\sT$ of the $J/\psi$ meson (the soft scale) is much smaller than its mass $M$ (the hard scale), namely $q_\sT \ll M$. Such an analysis could be performed at the future Electron-Ion Collider (EIC) planned in the United States~\cite{Boer:2011fh,Accardi:2012qut}.

In the present paper, we study instead a somewhat complementary process: the associated production of a $J/\psi$ or a $\Upsilon$ meson and a hadronic jet in  deep-inelastic electron-proton scattering, $e\, p \to e\, J/\psi \, (\Upsilon)\, {\rm jet } \, X$, where the initial proton can be either unpolarized or transversely polarized in the virtual photon-proton center of mass frame. In this case the soft scale is given by the total transverse momentum of the  $J/\psi $ + jet pair, required to be much smaller than its invariant mass, while the individual transverse momenta of the $J/\psi$ and the jet need not to be small. The advantage is that, by varying the invariant mass of the pair, one can now access a  range of scales, in contrast to the very narrow range around $M$ in $e\,p\to e\,J/\psi\, X$. In principle, this range of scales provides the opportunity to map out the TMD evolution, although its practical feasibility still remains to be seen.

Furthermore, by looking at the $J/\psi$ production in association with a jet, we probe a kinematic region in which the variable $z$, describing the energy fraction of the virtual photon transferred to the $J/\psi$ meson in the proton rest frame, assumes values smaller than one, where the background from diffraction events is negligible~\cite{Kniehl:2001tk}. This is in contrast with $e\,p\to e\,J/\psi\, X$: here one probes the endpoint region $z=1$, where diffraction events are expected to be present. Although such a background can be suppressed  by imposing that the virtuality of the photon is large enough~\cite{Bacchetta:2018ivt,Fleming:1997fq}, this will lead to a reduction of the nondiffractive signal as well.

The theoretical approach adopted in this analysis is the TMD formalism, as already mentioned,  in combination with nonrelativistic QCD (NRQCD), the effective field theory that provides a rigourous treatment of heavy-quarkonium production and decay~\cite{Bodwin:1994jh}. This formalism allows for a separation of short-distance coefficients, which can be calculated perturbatively in QCD, and long-distance matrix elements (LDMEs), which are of nonperturbative origin and are expected to scale  with a definite power of the heavy-quark velocity $v$ in the limit $v\ll 1$.  Hence, in addition to the common expansion in $\alpha_s$, NRQCD introduces a further expansion in $v$, with $v^2 \simeq 0.3$  for charmonium and $v^2 \simeq 0.1$ for bottomonium. As a consequence, one needs to take into account all the Fock states of $Q \overline Q$, the heavy quark-antiquark pairs produced in the hard scattering. Fock states are denoted by $^{2S+1}L_J^{(c)}$, where  $S$ is  the spin of the pair, $L$ the orbital angular momentum, $J$ the total angular momentum and $c$ the color configuration, with $c=1, 8$. For an $S$-wave quarkonium state like the $J/\psi$ and $\Upsilon$ mesons, the dominant contribution in the $v$ expansion, {\it i.e.}\ in the limit $v\to 0$, reduces to the traditional color-singlet model (CSM)~\cite{Berger:1980ni,Baier:1983va}, in which the heavy-quark pair is directly produced in a color-singlet (CS) state, with the same quantum numbers as the observed quarkonium. In addition, NRQCD predicts the existence of the color-octet (CO) mechanism, according to which the $Q\overline Q$ pair can be produced at short distances also in CO states with different angular momentum and spin and subsequently evolves into the physical CS quarkonia by the nonperturbative emission of soft gluons\footnote{For  an exhaustive overview on the phenomenology  of quarkonium production, see Ref.~\cite{Lansberg:2019adr} and references therein.}.

While in  $e\,p\to e\,J/\psi\, X$ at small $q_\sT$ the CO mechanism should be the dominant one~\cite{Bacchetta:2018ivt}, for the process under study both the CS and CO contributions to the cross section have to be taken into account.

Coming back to the process dependence of the gluon TMDs,  in the reaction under study the gauge links are future pointing. Hence, at small $x$ they correspond to the Weisz\"acker-Williams (WW) gluon distributions~\cite{Dominguez:2011br}.
As pointed out in Ref.~\cite{Boer:2015pni}, the WW gluon TMDs for a transversely polarized proton are suppressed by a factor of $x$ with respect to the unpolarized WW gluon TMDs. One therefore expects that the transverse spin asymmetries presented here could be reduced in the small-$x$ limit. This is a property that could be tested at the EIC, since there one  could probe both small and large $x$ regions. Furthermore, we point out that the gluon TMDs probed in $e\, p \to e\, J/\psi \, {\rm jet } \, X$   and in  $e\, p \to e\, J/\psi  \, X$ can be directly related, pending factorization theorems\footnote{See for example Ref. \cite{Li:2013mia} for a discussion about the soft factor.}, to the ones characterized by past-pointing gauge links which are also of the WW type, appearing for example in Higgs or heavy scalar quarkonium production in proton-proton collisions. This in principle allows to cross check the results obtained at the EIC and LHC for unpolarized protons, and, in case a polarized fixed target experiment will be achieved at the LHC~\cite{Brodsky:2012vg,Hadjidakis:2018ifr,Aidala:2019pit}, for transversely polarized protons as well.

The paper is organized as follows: in Sec.~\ref{calculation} we present the theoretical approach adopted and all details of the calculation, including kinematics and its region of validity. In Sec.~\ref{analytic} we give the analytic expressions of our results within NRQCD. In Sec.~\ref{sec:HQ} we compute the azimuthal modulations of interest in order to extract the TMDs under study. In Sec.~\ref{results} we show the corresponding numerical results adopting both known positivity bounds for the gluon TMDs and the McLerran-Venugopalan model. Finally, in Sec.~\ref{conclusions} we collect our conclusions and final remarks.

\section{Outline of the calculation}
\label{calculation}
\begin{figure}[t]
\begin{center}
\includegraphics[scale=.45]{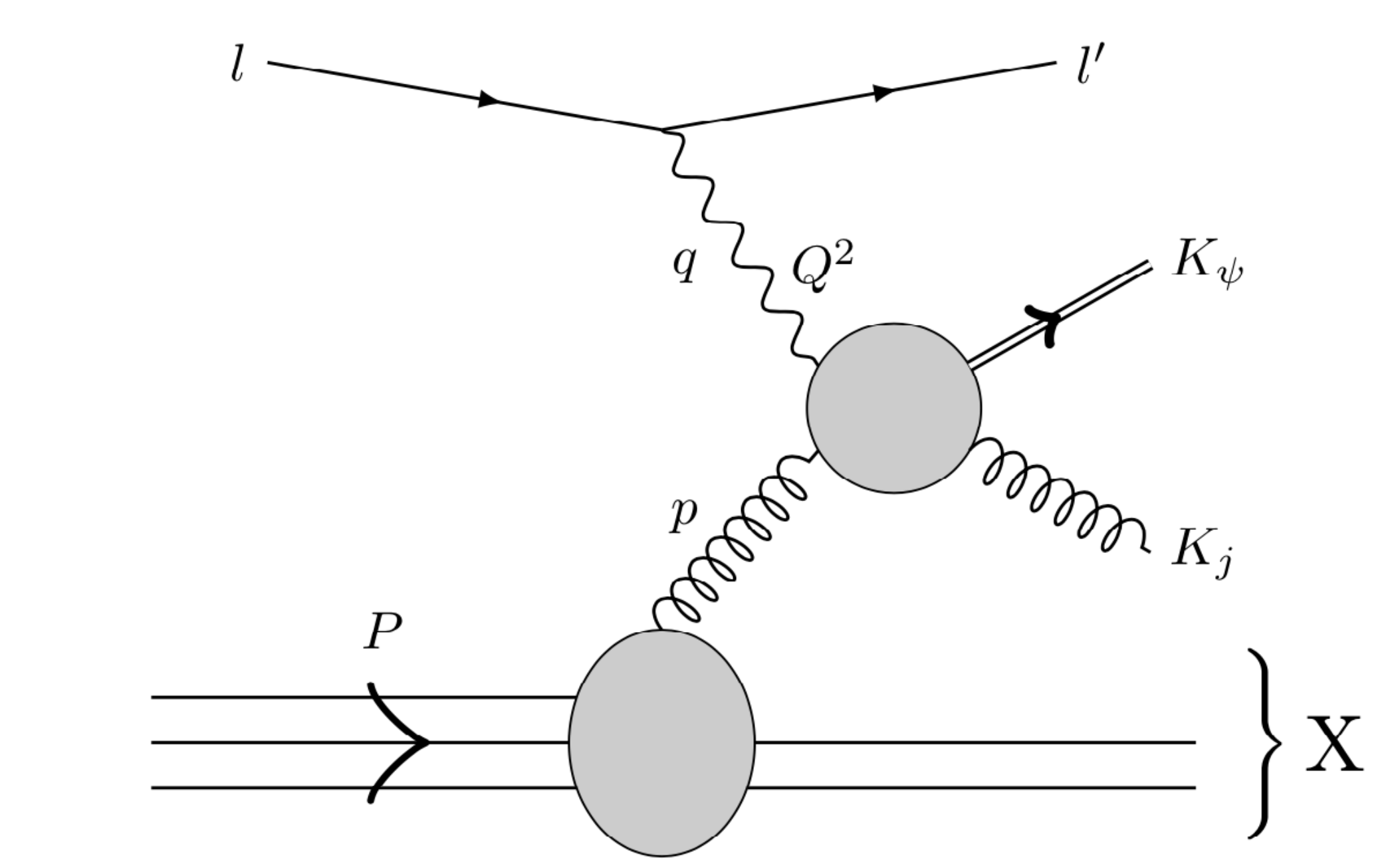}
\end{center}
\caption{Schematic illustration of the reaction $e(\ell) + p(P,S) \to e(\ell^{\prime}) + J/\psi(K_\psi) +  \text{jet}(K_{\rm j})+ X$ with the four-momentum assignements, where $X$ is the proton remnant and $Q^2 \equiv -q^2$.}
\label{fig:kinematics}
\end{figure}
We consider the process depicted in Fig. \ref{fig:kinematics},
\begin{equation}
e(\ell) + p(P,S) \to e(\ell^{\prime}) + J/\psi(K_\psi) +  \text{jet}(K_{\rm j})+ X\,,
\end{equation}
where the proton is polarized with polarization vector $S$, while we  do not observe the polarization of the other particles. The kinematics is defined with the help of a Sudakov decomposition in terms of two lightlike vectors, here chosen to be the momentum $P$ of the incoming proton, and a second vector $n$ which fulfills the relations  $n\cdot P=1$ and $n^{2}=0$. In a frame where the three-momenta of the incoming proton and the virtual photon exchanged in the reaction lie on the $\hat z$-axis, one has the following expressions for the momenta of the incoming parton ($p$), virtual photon ($q$),
outgoing jet ($K_{j}$) and quarkonium ($K_{\psi}$):
\begin{equation}
\begin{aligned}p^\mu & =xP^\mu+  (p\cdot P -M^2 x) n^\mu + p_{\sT}^\mu \approx x P^\mu + p^\mu_\sT\,,\\
q^\mu & =-\xB P^\mu+(P\cdot q)n^\mu\,,\\
K_{j}^\mu & =\frac{\bm{K}_{j\perp}^{2}}{2(1-z)P\cdot q}P^\mu+(1-z)(P\cdot q)n^\mu+K_{j\perp}^\mu\,,\\
K_{\psi}^\mu & =\frac{M^{2}+\bm{K}_{\psi\perp}^{2}}{2zP\cdot q}P^\mu+z(P\cdot q)n^\mu +K_{\psi\perp}^\mu\,.
\end{aligned}
\label{eq:Sudakovmomenta}
\end{equation}
In the above, $x = p \cdot n$, $M$ is the mass of the quarkonium, while the virtuality of the photon is defined as $Q^2=-q^2$, the Bjorken-$x$ variable is given by $\xB =Q^{2}/2P\cdot q$, and  $z=P\cdot K_{\psi}/P\cdot q$ is the energy fraction transferred from the photon to the quarkonium in the proton rest frame.  We clarify that, in our conventions, the subscript {\Large{$\sT$}} refers to a  soft transverse momentum, whereas $\perp$ denotes the large transverse component of a measurable hadronic momentum. In this process, and for our specific choice of the reference frame, the two directions are the same, as can be seen from the two-dimensional delta function in Eq.~\eqref{eq:delta} below.
Moreover, the momentum $\ell$ of the incoming lepton reads:
\begin{equation}
\ell^\mu=\frac{1-y}{y}\xB P^\mu+\frac{1}{y}(P\cdot q)n^\mu+\frac{\sqrt{1-y}}{y}\, Q\, \hat{\ell}^\mu_{\perp}\,,
\end{equation}
with $y=P\cdot q/P\cdot \ell$ the inelasticity variable, and from which
the momentum $\ell'$ of the scattered lepton can be obtained as $\ell'=\ell-q$. Similarly, for the proton spin vector
\begin{align}
S^\mu = \frac{S_L}{M_p}\, \left (  P^\mu - \frac{M_p^2}{P\cdot n} \, n^\mu \right ) + S^\mu_\sT\,,
\end{align}
with $S_L^2 + \bm S_\sT^2 = 1$ and $M_p$ denotes the proton mass.

Assuming TMD factorization, the cross section can be written as~\cite{Pisano:2013cya}:
\begin{equation}
\begin{aligned}\mathrm{d}\sigma & =\frac{1}{2s}\frac{\mathrm{d}^{3}\ell'}{(2\pi)^{3}2E_{\ell'}}\frac{\mathrm{d}^{3}K_{\psi}}{(2\pi)^{3}2E_{\psi}}\frac{\mathrm{d}^{3}K_{j}}{(2\pi)^{3}2E_{j}}\int\mathrm{d}x \, \mathrm{d}^{2}\bm{p}_{\sT}(2\pi)^{4}\delta^{4}\bigl(q+p-K_{j}-K_{\psi}\bigr)\\
 &\qquad \qquad  \times\frac{1}{Q^{4}}L^{\mu\nu}(\ell,q)\, \Gamma^{\rho\sigma}_{g}(x,\bm{p}_\sT) \, H_{\mu\rho}\, H_{\nu\sigma}^*\;.
\end{aligned}
\label{eq:csTMD}
\end{equation}
In the above expression, $L^{\mu\nu}$ is the usual leptonic tensor, defined as:
\begin{equation}
\begin{aligned}
L^{\mu\nu}(\ell,q) &= \frac{g_e^2}{2} \mathrm{tr} (\slashed{\ell}\gamma^\mu \slashed{\ell}' \gamma^{\nu})=g_e^2\left(-Q^2 g^{\mu\nu}+2 \ell^{\{\mu} \ell'^{\nu\}}\right )\,,
\end{aligned}
\label{eq:x}
\end{equation}
where the factor $1/2$ takes care of the spin average of the incoming electron, and where we introduced the symmetrization operator $p^{ \{ \mu} q^{\nu \} }=p^\mu q^\nu + p^\nu q^\mu$.
The  function  $H$ in Eq.~(\ref{eq:csTMD}) is the scattering amplitude for the dominant  partonic subprocess $\gamma^{*}(q) + g(p) \to  J/\psi(K_\psi) +  g(K_{\rm j})$. As such, it  takes into account also the bounding mechanism into the $J/\psi$.  Moreover, it is understood that it includes the factor  $1/(N_c^2-1)$ coming from the average over the incoming gluon colors.

Finally, $\Gamma^{\mu\nu}_g$ in Eq.~(\ref{eq:csTMD}) is the gluon correlator, encoding the gluon content of the proton,  which can be parameterized in terms of gluon TMDs~\cite{Mulders:2000sh,Meissner:2007rx,Boer:2016xqr}. For an unpolarized proton, omitting the gauge link dependence, this correlator is given by:
\begin{align}
 {\Gamma}_U^{\mu\nu}(x,\bm p_\sT )  = & \frac{1}{2x}\,\bigg \{-g_\sT^{\mu\nu}\,f_1^g (x,\bm p_\sT^2) +\bigg(\frac{p_\sT^\mu p_\sT^\nu}{M_p^2}\,
    {+}\,g_\sT^{\mu\nu}\frac{\bm p_\sT^2}{2M_p^2}\bigg) \,h_1^{\perp\,g} (x,\bm p_\sT^2) \bigg \} \,,
\label{eq:PhiparU}
\end{align}
where $g_\sT^{\mu\nu} \equiv g^{\mu\nu} - P^{\{\mu} n^{\nu\}}$. In the above expression, $f_1^g (x,\bm p_\sT^2)$ is the TMD unpolarized gluon distribution, while $h_1^{\perp\,g} (x,\bm p_\sT^2) $ is the distribution of linearly polarized gluons in an unpolarized proton. Both distributions are $T$-even, implying that they can be nonzero even in processes where neither initial nor final state interactions are present.
In the case of a transversely polarized proton, its correlator can be parameterized as:
\begin{align}
 {\Gamma}_T^{\mu\nu}(x,\bm p_\sT )   = & \frac{1}{2x}\,\bigg \{g^{\mu\nu}_\sT\,
    \frac{ \epsilon^{\rho\sigma}_\sT p_{\sT \rho}\, S_{\sT\sigma}}{M_p}\, f_{1T}^{\perp\,g}(x, \bm p_\sT^2) + i \epsilon_\sT^{\mu\nu}\,
    \frac{p_\sT \cdot S_\sT}{M_p}\, g_{1T}^{g}(x, \bm p_\sT^2) \nonumber \\
    &  \,  + \,  \frac{p_{\sT \rho}\,\epsilon_\sT^{\rho \{ \mu}p_\sT^{\nu \}}}{2M_p^2}\,\frac{p_\sT\cdot S_\sT }{M_p} \, h_{1 T}^{\perp\,g}(x, \bm p_\sT^2)\,- \,\frac{p_{\sT \rho} \epsilon_\sT^{\rho \{ \mu}S_\sT^{\nu \}}\, + \,
      S_{\sT\rho} \epsilon_\sT^{\rho \{ \mu } p_\sT^{\nu \}}}{4M_p} \, h_{1T}^{g}(x, \bm p_\sT^2) \,\,\bigg \}\,,
\label{eq:PhiparT}
\end{align}
where we have introduced the antisymmetric transverse projector $\epsilon_\sT^{\mu\nu} = \epsilon^{\mu\nu\alpha\beta} P_\alpha n_\beta$, with $\epsilon_\sT^{12} = +1$.
In the symmetric part of the correlator, $\Gamma_T^{\{\mu\nu\}}/2$,  three $T$-odd TMDs appear: the gluon Sivers function $f_{1T}^{\perp\,g}(x, \bm p_\sT^2)$, as well as the TMDs $h_{1T}^{\perp g}$  and $h_{1T}^{g}$ which are chiral-even distributions of linearly polarized gluons inside a transversely polarized proton.
In analogy with the transversity TMD for quarks, one introduces the combination
\begin{equation}
h_1^g \equiv h_{1T}^g +\frac{\bm p_\sT^2}{2 M_p^2}\,  h_{1T}^{\perp\,g}\,,
\label{eq:h1}
\end{equation}
which vanishes upon integration over transverse momentum~\cite{Boer:2016fqd}, in contrast to its quark counterpart. Finally, the distribution $ g_{1T}^{g}(x, \bm p_\sT^2)$ is $T$-even and represents the circular polarized gluon content of a transversely polarized proton. Note that the factor $1/2$, necessary to average over the incoming gluon polarization, is already taken into account in Eqs.~(\ref{eq:PhiparU}) and (\ref{eq:PhiparT}).

Integrating out the azimuthal angle of the final lepton $\ell'$~\cite{Graudenz:1993tg}, one has
\begin{equation}
\begin{aligned}\frac{\mathrm{d}^{3}\ell'}{(2\pi)^{3}2E_{\ell'}} & =\frac{\mathrm{d}Q^{2}\mathrm{d}y}{16\pi^{2}}\;.\end{aligned}
\end{equation}
Moreover, we can write:
\begin{equation}
\begin{aligned}\frac{\mathrm{d}^{3}K_{\psi}}{(2\pi)^{3}2E_{\psi}} & =\frac{\mathrm{d}z\mathrm{d}^{2}\bm{K}_{\psi\perp}}{(2\pi)^{3}2z}\;,\quad\frac{\mathrm{d}^{3}K_{j}}{(2\pi)^{3}2E_{j}}=\frac{\mathrm{d}\bar{z}\mathrm{d}^{2}\bm{K}_{j\perp}}{(2\pi)^{3}2\bar{z}}\;,\end{aligned}
\end{equation}
and
\begin{equation}
\begin{aligned} & \delta^{4}\bigl(q+p-K_{j}-K_{\psi}\bigr)\\
 & =\frac{2}{ys}\delta\bigl(1-z-\bar{z}\bigr)\delta\left(x-\frac{\bar{z}(M^2+\bm{K}_{\psi\perp}^2)+z \bm{K}_{j\perp}^2 + z \bar{z}Q^2}{z(1-z)ys}\right )\delta^{2}\bigl(\bm{p}_{\sT}-\bm{K}_{j\perp}-\bm{K}_{\psi\perp}\bigr)\;,
\label{eq:delta}
\end{aligned}
\end{equation}
where we made use of the relation $Q^{2}=\xB  ys$, with $s=(\ell+P)^2$ the square of the center-of-mass energy. Finally, integrating over $\bar{z}$, $\bm{p}_{_T}$ and $x$, the cross section in Eq.~(\ref{eq:csTMD}) can be rewritten as
\begin{equation}
\begin{aligned}\frac{\mathrm{d}\sigma}{\mathrm{d}z\,\mathrm{d}y\,\mathrm{d}\xB \,\mathrm{d}^{2}\bm{q}_{\sT}\mathrm{d}^{2}\bm{K}_{\perp}} & =\frac{1}{(2\pi)^{4}}\frac{1}{16sz(1-z)Q^{4}}L^{\mu\nu}(\ell,q)\,  \Gamma^{\rho\sigma}_{g}(x,\bm{q}_{\sT})\, H _{\mu\rho}\, H^*_{\nu\sigma}\;.\end{aligned}
\label{eq:csdetail}
\end{equation}

In the above expression,
\begin{equation}
\begin{aligned}\bm{q}_{\sT} & \equiv \bm{K}_{\psi\perp}+\bm{K}_{j\perp}\,,\quad \bm{K}_{\perp}\equiv\frac{\bm{K}_{\psi\perp}-\bm{K}_{j\perp}}{2}\,.\end{aligned}
\end{equation}
The region of validity of our calculation is then the one where there are two strongly ordered scales: $q_\sT \equiv |\bm{q}_{\sT} | \ll K_\perp \equiv |\bm{K}_{\perp}|$, corresponding to the setup where the $J/\psi$ and the jet are produced almost back to back in the plane transverse to the collision axis. We can then set $\bm{K}_{\perp}\simeq \bm{K}_{\psi \perp} \simeq -\bm{K}_{j\perp}$ in our calculation, and retain only those terms which are, at most, of the order of $ q_\sT /M_p$, discarding the ones suppressed by powers of $q_\sT/M$ or $q_\sT/K_\perp$.

Note that, integrating Eq.~(\ref{eq:csdetail}) over $q_\sT$, one recovers the collinear result~\cite{Kniehl:2001tk}
\begin{equation}
\begin{aligned}\frac{\mathrm{d}\sigma}{\mathrm{d}y \,\d Q^2}  =\frac{1}{(2\pi)^{3}}\int^1_{x_{\rm min}} \d x\,\int^0_{\hat t_{\rm min}}{\mathrm{d}\hat{t}}\, \frac{1}{64 x^2 y^2 s^2 Q^{4}}L^{\mu\nu}(-g^{\rho\sigma}_\sT ) f^g_1(x) \, H_{\mu\rho}\, H^*_{\nu\sigma}\,,\end{aligned}
\end{equation}
where we have introduced the usual Mandelstam variables $\hat s$ and $ \hat t$ for the subprocess $\gamma^*g \to J/\psi\, g$, and where   $x_{\rm min}= (Q^2+M^2)/{(y s)}$,  $\hat t_{\rm min} = -(\hat s + Q^2)(\hat s -M^2)/\hat s$.
\begin{figure}[t]
\begin{center}
\includegraphics[scale=.35]{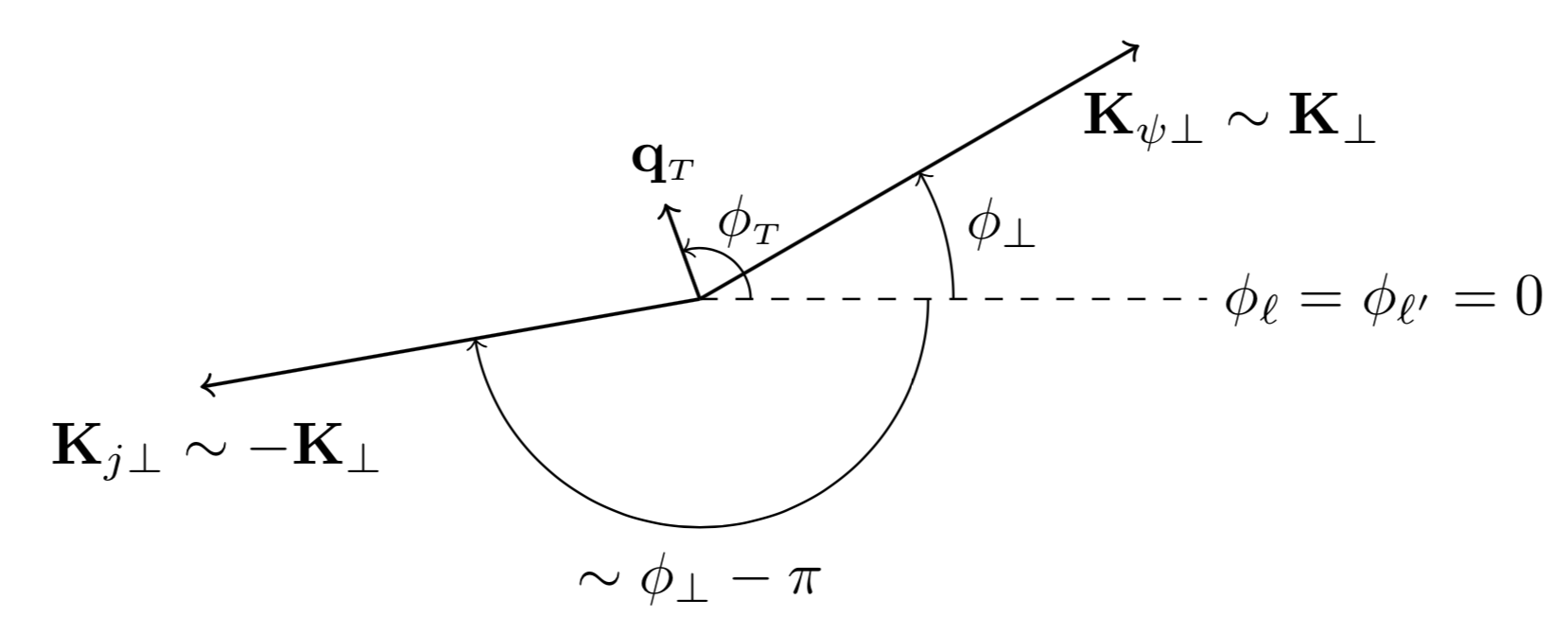}
\end{center}
\caption{Azimuthal angles for the process $e\, p \to  e\, J/\psi\, {\rm jet}\, X$ in a reference frame where $\phi_\ell = \phi_{\ell^\prime} = 0$. The vectors $\bm{q}_{\sT}  = \bm{K}_{\psi\perp} + \bm{K}_{j\perp}$ and  $\bm{K}_{\perp} = (\bm{K}_{\psi\perp} - \bm{K}_{j\perp})/{2}$ define the azimuthal angles $\phi_\sT$ and $\phi_\perp$, respectively.}
\label{fig:angles}
\end{figure}

\section{Angular structure of the cross section: analytic results}
\label{analytic}
In the reference frame defined above, and measuring the azimuthal angles  w.r.t. the lepton plane ($\phi_{\ell}=\phi_{\ell^\prime}=0)$ as depicted in Fig.~\ref{fig:angles}, we  denote by $\phi_S$, $\phi_\sT$ and $\phi_\perp$ the azimuthal angles of the three-vectors $\bm S_\sT$,  $\bm q_\sT$ and  $\bm K_\perp$, respectively. In the region $q_\sT \ll K_\perp$,  we obtain that the differential cross section can be written in the form~\cite{Pisano:2013cya}
\begin{equation}
\frac{\d\sigma}
{\d z\,\d y\,\d\xB \,\d^2\bm{q}_{\sT} \d^2\bm{K}_{\perp}} \equiv \d\sigma (\phi_S, \phi_\sT,\phi_\perp) =    \d\sigma^U(\phi_\sT,\phi_\perp)  +  \d\sigma^T (\phi_S, \phi_\sT,\phi_\perp)  \,.
\label{eq:cs}
\end{equation}
At leading order (LO) in perturbative QCD, we find
\begin{align}
\d\sigma^{U} & ={\cal N}\,\bigg[\bigl(\mathcal{A}_{0}^{eg}+\mathcal{A}_{1}^{eg} \cos\phi_{\perp}+\mathcal{A}_{2}^{eg} \cos2\phi_{\perp}\bigr)f_{1}^{g}(x,\bm{q}_{\sT}^{2})+\bigl(\mathcal{B}_{0}^{eg} \cos2\phi_{\sT}+\mathcal{B}_{1}^{eg} \cos(2\phi_{\sT}-\phi_{\perp})\nonumber\\
 & \qquad\quad+\mathcal{B}_{2}^{eg} \cos2(\phi_{\sT}-\phi_{\perp})+\mathcal{B}_{3}^{eg} \cos(2\phi_{\sT}-3\phi_{\perp})+\mathcal{B}_{4}^{eg} \cos(2\phi_{\sT}-4\phi_{\perp})\bigr)\frac{ \bm q_\sT^2  }{M_p^2} \,h_{1}^{\perp\, g} (x,\bm{q}_{\sT}^2)\bigg]\,,
\label{eq:csU}
\end{align}
while
\begin{align}
\d\sigma^T
  & =  {{\cal N}\,\vert \bm S_\sT\vert}\, \bigg[\sin(\phi_S -\phi_\sT) \bigl( \mathcal{A}^{eg}_0 + \mathcal{A}^{eg}_1 \cos \phi_\perp  + \mathcal{A}^{eg}_2 \cos 2 \phi_\perp \bigr ) \frac{ |\bm q_\sT | }{M_p}f_{1T}^{\perp\, g} (x,\bm{q}_{\sT}^2)\nonumber \\
 & + \cos(\phi_S-\phi_\sT) \bigl( \mathcal{B}^{eg}_0 \sin 2 \phi_\sT  +  \mathcal{B}^{eg}_1 \sin(2\phi_\sT-\phi_\perp) + \mathcal{B}^{eg}_2 \sin 2 (\phi_\sT-\phi_\perp) \nonumber \\
 &\quad+ \mathcal{B}^{eg}_3 \sin( 2\phi_\sT-3\phi_\perp)  + \mathcal{B}^{eg}_4\sin(2\phi_\sT- 4\phi_\perp)  \bigr) \frac{ |\bm q_\sT |^3  }{M_p^3} \,h_{1T}^{\perp\, g} (x,\bm{q}_{\sT}^2) \nonumber\\
& + \bigl (\mathcal{B}_0^{eg} \sin(\phi_S+\phi_\sT) + \mathcal{B}_1^{eg} \sin(\phi_S + \phi_T-\phi_\perp) + \mathcal{B}_2^{eg} \sin(\phi_S+\phi_\sT-2\phi_\perp) \nonumber \\
&  \quad  +  \mathcal{B}_3^{eg} \sin(\phi_S+\phi_\sT-3\phi_\perp) + \mathcal{B}_4^{eg} \sin(\phi_S+\phi_\sT-4 \phi_\perp) \bigr)\frac{ |\bm q_\sT | }{M_p} h_{1T}^{g} (x,\bm{q}_{\sT}^2) \bigg ]\, .
\label{eq:csT}
\end{align}
The normalization factor $\cal N$ is given by
\begin{equation}
\begin{aligned}
{\cal N}&=\frac{8\alpha^2\alpha_s^2e_Q^2}{3yMQ^2M_\perp^{10}D}\,,
\end{aligned}
\end{equation}
where $e_Q$ is the electric charge of the heavy quark in units of the proton charge, and
\begin{equation}
D\equiv D \left (z, \frac{Q^2}{M_\perp^2} , \frac{M^2}{M_\perp^2}\right ) = \left [1 - z (2-z) \frac{M^2}{M_\perp^2} \right ]^2\left [1 + z (2-z) \frac{Q^2}{M_\perp^2} \right ]^2 \left [1-z \left ( \frac{M^2}{M_\perp^2} - (1-z)\frac{Q^2}{M_\perp^2} \right ) \right ]\,,
\label{eq:Den}
\end{equation}
with the transverse mass $M_\perp$ defined as $M_\perp=\sqrt{M^2+\bm{K}_\perp^2}$.

The amplitudes of the modulations ${\cal A}_l^{eg}$, with $l = 0,1,2$, describe the interaction of an electron with an unpolarized gluon, through the exchange of a photon which can be in different polarization states~\cite{Pisano:2013cya}, and can be expressed as
\begin{align}
{\cal A}_0^{eg} & =  [1+(1-y)^2]\,{\cal A}_{U+L}^{\gamma^*g} -y^2 \,{\cal A}_{L}^{\gamma^*g}\,,\nonumber \\
{\cal A}_{1}^{eg} & = (2-y) \sqrt{1-y}\,{\cal A}_{I}^{\gamma^*g}\,,\nonumber \\
 {\cal A}_{2}^{eg} & =  2 (1-y)\, {\cal A}_{T}^{\gamma^*g}\,,
\label{eq:Agstar}
\end{align}
where the subscripts $i = U+L$, $L$, $I$, $T$ refer to the specific polarizations of the photon: unpolarized plus longitudinal, longitudinal, interference transverse-longitudinal and transverse, respectively~\cite{Pisano:2013cya,Brodkorb:1994de}. Namely, denoting by ${\cal A}^{\gamma^*g}_{\lambda_\gamma,\lambda_\gamma^\prime}$,  with $\lambda_\gamma, \lambda_\gamma^\prime =0, \pm 1$,
the  helicity amplitudes squared for the process $\gamma^* g \to Q \overline Q \big [^{2 S +1}L^{(c)}_J \big ]\,g$,  the following relations are fulfiled
\begin{eqnarray}
{\cal A}_{U+L}^{\gamma^*g}&  \propto & {\cal A}_{++} ^{\gamma^*g}+ {\cal A}_{--}^{\gamma^*g} + {\cal A}_{00}^{\gamma^*g}\, ,
\nonumber \\
{\cal A}_{L}^{\gamma^*g} &  \propto & {\cal A}^{\gamma^*g}_{00}\,, \nonumber \\
{\cal A}_{I} ^{\gamma^*g}&  \propto & {\cal A}^{\gamma^*g}_{0+} +{\cal A}^{\gamma^*g}_{+0}-{\cal A}_{0-}^{\gamma^*g} -{\cal A}_{-0}^{\gamma^*g} \,, \nonumber \\
{\cal A}_{T}^{\gamma^*g} &  \propto & {\cal A}_{+-}^{\gamma^*g} + {\cal A}_{-+}^{\gamma^*g}~\,,
\label{eq:Ahel}
\end{eqnarray}
where we have omitted numerical prefactors.
Analogously,  for the amplitudes $B_m^{eg}$ with $m=0,1,2,3,4$, one can write
\begin{align}
{\cal B}_2^{e g} & =  [1+(1-y)^2]\,{\cal B}_{U+L}^{\gamma^*g} -y^2 \,{\cal B}_{L}^{\gamma^*g}\,,\nonumber \\
{\cal B}_{m}^{e g} & = (2-y) \sqrt{1-y}\,{\cal B}_{m I}^{\gamma^*g }\qquad (m =1,3) \,,\nonumber \\
 {\cal B}_{m}^{e g} & =  2 (1-y)\, {\cal B}_{m T}^{\gamma^*g}\qquad (m = 0,4)\,.
\label{eq:Agstar}
\end{align}

\begin{figure}[t]
\begin{center}
\includegraphics[scale=.45]{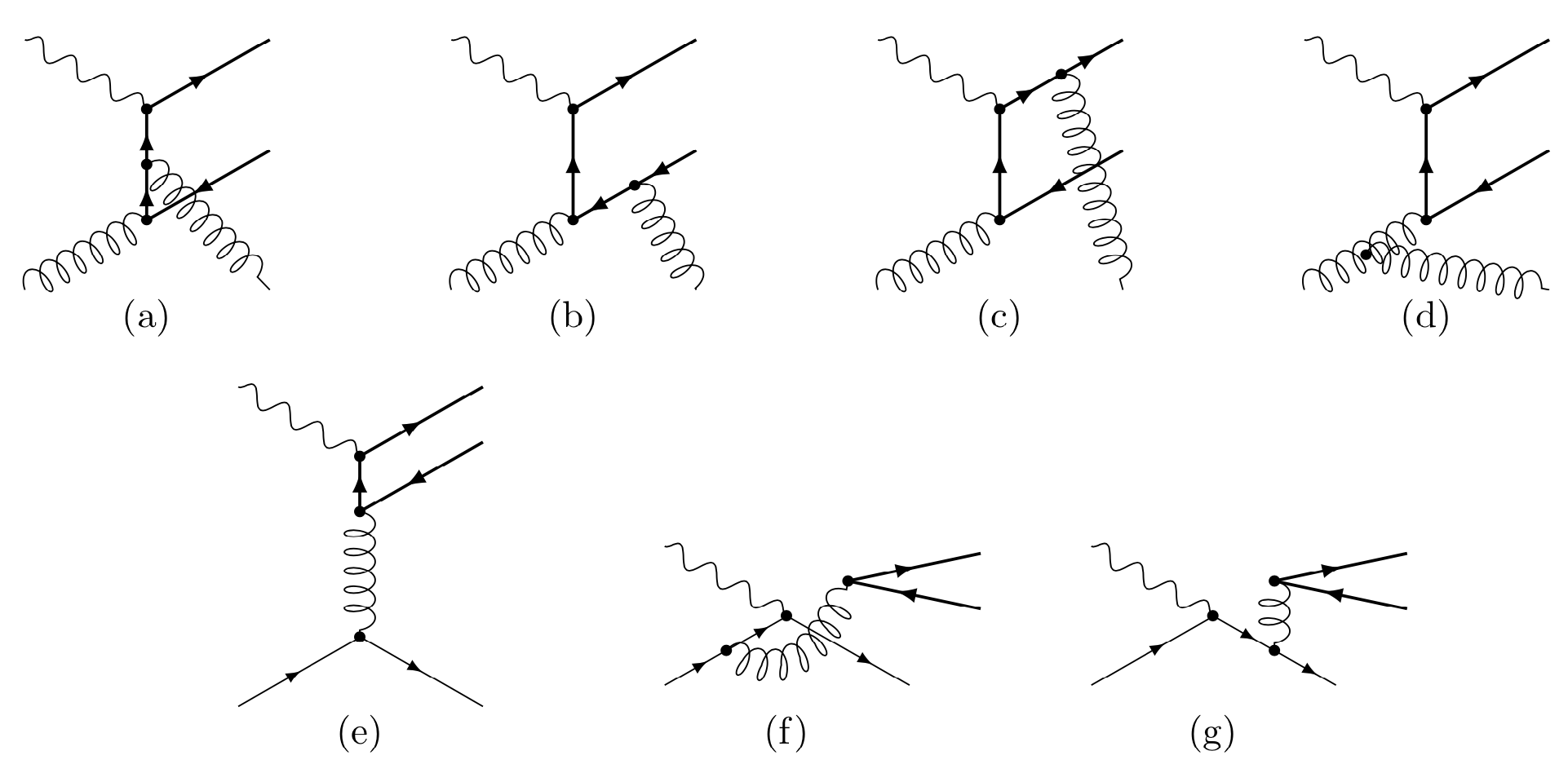}
\end{center}
\caption{Feynman diagrams representing the partonic subprocesses underlying the reaction $e \, p \to  e \, J/\psi \,  \mathrm{jet} \, X$; (a)-(d):  $\gamma^* g \to Q \overline Q[^{2S+1} L_J^{(c)}]\, g$,  (e)-(g):  $\gamma^* q \to Q \overline Q[^{2S+1}L_J^{(c)}]\, q$. The crossed diagrams of (a)-(e), obtained by  reversing the heavy quark lines, are included in the calculations but not shown explicitly.  Only diagrams (a)-(c) contribute to the CS production mechanism.}
\label{fig:subp}
\end{figure}

The amplitudes ${\cal{A}}^{\gamma^*g}$ and ${\cal{B}}^{\gamma^*g}$ have been calculated at LO in the framework of NRQCD, taking into account both the CS and CO contributions corresponding to  the Feynman diagrams in Fig.~\ref{fig:subp}.

According to the CS mechanism, the heavy-quark pair is produced in the hard process $\gamma^*g \to Q\overline{Q}[^{2S+1}L_J^{(c)}]\, g$  directly with the quantum numbers of the observed quarkonium, which in the case of the $J/\psi$  meson are $S=1,\;L=0$ and $J=1$. This means that the pair is in a $^3S_1^{(1)}$ state. On the other hand, the  allowed CO states are $^1S_0^{(8)}$, $^3S_1^{(8)}$, $^3P_{J}^{(8)}$, with $J = 0,1,2$. Moreover, the quark-induced subprocess  $\gamma^*q \to Q\overline{Q}[^{2S+1}L_J^{(8)}]\, q$  should also be taken into account, in principle. However,  since it turns out to be negligible in the kinematic region under study (see Sec.~\ref{results}), we will not consider it in the following.

In the calculation of ${ \cal A}_i^{\gamma^*g}$ and ${ \cal B}_i^{\gamma^*g}$, all the CS and CO perturbative contributions need to be added, each one weighted by its corresponding LDME, denoted by $\langle 0 \vert{ \cal O}_c (^{2S+1}L_J) \vert 0\rangle$, encoding the soft hadronization of a $Q \overline Q $ pair with angular momentum configuration  $^{2S+1}L_J$ and color assignment $c=1,8$, into a colorless state with the quantum numbers of the $J/\psi$. One can write
\begin{align}
\label{eq:ABNRQCD}
{\cal A}_i^{\gamma^*g} & ={\cal C}_{CS}\, \langle 0 \vert { \cal O}_1 (^3S_1)   \vert 0\rangle  \,{\cal A}_i^{CS}  + {\cal C}_8^{^1S_0}\,\langle 0 \vert{ \cal O}_8 (^1S_0) \vert 0\rangle\,{\cal A}_i^{^1S_0}\,+{\cal C}_8^{^3S_1}\,\langle 0 \vert { \cal O}_8 (^3S_1) \vert 0\rangle\,{\cal A}_i^{^3S_1}+{\cal C}_8^{^3P_0}\,\langle 0 \vert { \cal O}_8 (^3P_0) \vert 0\rangle\,{\cal A}_i^{^3P_0}\nonumber\\
&\qquad \qquad +{\cal C}_8^{^3P_1}\,\langle 0 \vert { \cal O}_8 (^3P_1) \vert 0\rangle\,{\cal A}_i^{^3P_1}+{\cal C}_8^{^3P_2}\,\langle 0 \vert { \cal O}_8 (^3P_2) \vert 0\rangle\,{\cal A}_i^{^3P_2}\,,
\end{align}
with an analogous expression holding for the amplitudes ${\cal B}_i^{\gamma^* g}$.

Making use of the heavy-quark spin symmetry relation~\cite{Bodwin:1994jh}
\begin{align}
\langle 0 \vert { \cal O}_8 (^3P_J) \vert 0\rangle  =(2J+1)\langle 0 \vert { \cal O}_8 (^3P_0) \vert 0\rangle\,,
\label{eq:HQSS}
\end{align}
which is valid up to higher order corrections in the small expansion parameter $v$,
and the fact that the color factors read
\begin{align}
{\cal C}_{CS} & = \frac{1}{N_c}\,, \nonumber \\
{\cal C}_1& \equiv{\cal C}_8^{^3S_1}=\frac{N_c^2-4}{2N_c}\,,\nonumber\\
{\cal C}_2& \equiv{\cal C}_8^{^1S_0}={\cal C}_8^{^3P_J}=\frac{N_c}{2}\quad\mathrm{for}\quad J=1,2,3\,,
\label{eq:ABNRQCD2}
\end{align}
Eq.~(\ref{eq:ABNRQCD}) can be simplified further to yield
\begin{align}
{\cal A}_i^{\gamma^*g}  &={\cal C}_{CS}\, \langle 0 \vert { \cal O}_1 (^3S_1) \vert 0\rangle\,{\cal A}_i^{CS} + {\cal C}_1\,\langle 0 \vert { \cal O}_8 (^3S_1) \vert 0\rangle\,{\cal A}_i^{^3S_1}\,+{\cal C}_2 \big( \langle 0 \vert { \cal O}_8 (^1S_0) \vert 0\rangle\,{\cal A}_i^{^1S_0}+\langle 0 \vert { \cal O}_8 (^3P_0) \vert 0\rangle\,{\cal A}_i^{^3P_J}\big)\,,\nonumber\\
{\cal B}_i^{\gamma^*g}  &={\cal C}_{CS}\, \langle 0 \vert { \cal O}_1 (^3S_1) \vert 0\rangle \, {\cal B}_i^{CS} + {\cal C}_1\,\langle 0 \vert { \cal O}_8 (^3S_1) \vert 0\rangle\,{\cal B}_i^{^3S_1}\,+{\cal C}_2 \big( \langle 0 \vert { \cal O}_8 (^1S_0) \vert 0\rangle\,{\cal B}_i^{^1S_0}+\langle 0 \vert { \cal O}_8 (^3P_0) \vert 0\rangle\,{\cal B}_i^{^3P_J}\big)\,,
\label{eq:ABNRQCD3}
\end{align}
where we defined ${\cal A}_i^{^3P_J}\equiv{\cal A}_i^{^3P_0}+3{\cal A}_i^{^3P_1}+5{\cal A}_i^{^3P_2}$, and ${\cal B}_i^{^3P_J}\equiv{\cal B}_i^{^3P_0}+3{\cal B}_i^{^3P_1}+5{\cal B}_i^{^3P_2}$.

The explicit  expressions for the hard scattering functions ${\cal{A}}_i^{\gamma^*g}$ and ${\cal{B}}_i^{\gamma^*g}$  we introduced above, depend on the production mechanism of the quarkonium we consider.
For the CS hard parts, we obtain:
\begin{align}
{\cal A}_{U+L}^{CS}
&=2\Big\{M_\perp^4 \big(M^2 (z^2-z+1)^2+Q^2 z^2\big)\nonumber\\
&+M_\perp^2 z \big(M^4 (-z^3+z-2)+2 M^2 Q^2 (2 z^5-8 z^4+12 z^3-7
   z^2-z+1)+Q^4 (z-1)^2 z\big)\nonumber\\
   &+M^2 z^2 \big(M^4 (z^2-2 z+2)+M^2 Q^2 (-4 z^4+18 z^3-32 z^2+26 z-7)\nonumber\\*
   &\qquad\qquad\qquad\qquad\qquad\qquad\qquad\qquad\qquad\qquad\qquad\qquad\qquad+Q^4 (z-1)^2
   (z^4-4 z^3+7 z^2-6 z+1)\big)\Big\}\,,\nonumber\\
  {\cal A}_{L}^{CS}
  &=2z^2Q^2 \Big\{ M_\perp^4+2 M^2 M_\perp^2 (2 z^4-8 z^3+12 z^2-8 z+1)-M^4 (3 z^4-12 z^3+18 z^2-12 z+2)\Big\}\,,\nonumber\\
    {\cal A}_{I}^{CS}
  &=4z^2(1-z)^2QK_\perp\Big\{M_\perp^2 \big(M^2 (2 z^2-4 z+3)+Q^2\big)- M^4 (2 z^2-4 z+3)\nonumber\\*
  &\qquad\qquad\qquad\qquad\qquad\qquad\qquad\qquad\qquad\qquad\qquad\qquad\qquad+M^2 Q^2 (2 z^4-8 z^3+12 z^2-8 z+1)\Big\}\,,\nonumber\\
   {\cal A}_{T}^{CS}
  &=2z^2(1-z)^2M^2K_\perp^2\big\{Q^2 (2 z^2-4 z+1)-M^2\big\}\,,
\end{align}
and
\begin{align}
{\cal B}_{U+L}^{CS}
&=z^2(1-z)^2M^2K_\perp^2 \big\{Q^2(6z^2-12z+5)-M^2\big\}\,,\nonumber\\
{\cal B}_{L}^{CS}
&=2z^2(1-z)^2M^2Q^2K_\perp^2(2z^2-4z+1)\,,\nonumber\\
{\cal B}_{0T}^{CS}
&=z^2 (1-z)^4 M^2 Q^2 \big\{Q^2(z-2)z-M^2\big\}\,,\nonumber\\
{\cal B}_{1I}^{CS}
&=2z^2 (1-z)^3 Q M^2 K_\perp \big\{Q^2(2z^2-4z+1)-M^2\big\}\,,\nonumber\\
{\cal B}_{3I}^{CS}
&=4z^2(1-z)^3 Q M^2 K_\perp^3\,,\nonumber\\
{\cal B}_{4T}^{CS}
&=z^2(1-z)^2M^2 K_\perp^4\,.
\end{align}

Since the CO expressions for the hard parts are very long and not very enlighting, we do not show them here. We note that our results for the angular independent part of the cross section fully agree with the ones published in Ref.~\cite{Kniehl:2001tk}, obtained within the framework of collinear NRQCD, while, to the best of our knowledge, the angular modulations are derived here for the first time.

\section{Azimuthal modulations}
\label{sec:HQ}

The cross sections in Eqs.~(\ref{eq:csU}) and (\ref{eq:csT}) exhibit various azimuthal modulations, and have the same structure as similar $2\to2$ processes such as $\gamma^* g \to Q \bar{Q} $ \cite{Pisano:2013cya}. These modulations can be exploited to extract ratios of specific gluon TMDs over the unpolarized one, $f_1^{ g}(x,\bm q_\sT^2 )$.
To this end, we define the following azimuthal moments
\begin{align}
A^{W(\phi_S,\phi_\sT)} & \equiv 2\,\frac {\int\d \phi_S\, \d \phi_\sT \,\d\phi_\perp\, W(\phi_S,\phi_\sT)\,\d\sigma(\phi_S,\,\phi_\sT,\,\phi_\perp)}{\int \d \phi_S\,\d\phi_\sT \,\d\phi_\perp\,\d\sigma(\phi_S,\phi_\sT,\phi_\perp)} \,,
\label{eq:mom}
\end{align}
where the denominator is given by
\begin{align}
\int \d\phi_S\,\d \phi_\sT \,\d\phi_\perp\,\d\sigma(\phi_S,\phi_\sT,\phi_\perp)  & =\int \d\phi_S\,\d \phi_\sT \,\d\phi_\perp\,\d\sigma^U(\phi_\sT,\phi_\perp)=  (2\pi)^3 {\cal N} {\cal A}^{eg}_0 f_{1}^{g}(x,\bm{q}_{\sT}^{2})\;.
\label{eq:A}
\end{align}

As an example, to extract the ratio $h_1^{\perp\, g}(x,\bm q_\sT^2 )/f_1^{ g}(x,\bm q_\sT^2 )$, where $h_1^{\perp\, g}$ is the distribution of linearly polarized gluons in an unpolarized proton, one could measure:
 \begin{align}
A^{\cos 2 \phi_\sT}&= \frac{\bm q_\sT^2}{M_p^2} \, \frac{ {\cal B}_0^{eg} }{{\cal A}_0^{e g}}\,\frac{h_1^{\perp\, g}(x,\bm q_\sT^2 )}{ f_1^{g}(x,\bm q_\sT^2 )}\nonumber\\
&= \frac{\bm q_\sT^2}{M_p^2} \, \frac{2(1-y) {\cal B}_{0T}^{\gamma^*g} }{ \big[1+(1-y)^2\big]\,{\cal A}_{U+L}^{\gamma^*g } -y^2 \,{\cal A}_{L}^{\gamma^*g}}\,\frac{h_1^{\perp\, g}(x,\bm q_\sT^2 )}{ f_1^{g}(x,\bm q_\sT^2 )}\,,
\label{eq:cos2phiT}
\end{align}
or
 \begin{align}
A^{\cos 2 (\phi_\sT-\phi_\perp)}&= \frac{\bm q_\sT^2}{M_p^2} \, \frac{{\cal B}_2^{eg} }{{\cal A}_0^{e g}}\,\frac{h_1^{\perp\, g}(x,\bm q_\sT^2 )}{ f_1^{g}(x,\bm q_\sT^2 )} \nonumber\\
&= \frac{\bm q_\sT^2}{M_p^2} \, \frac{\big[1+(1-y)^2\big]\,{\cal B}_{U+L}^{\gamma^*g} -y^2 \,{\cal B}_{L}^{\gamma^*g}}{ \big[1+(1-y)^2\big]\,{\cal A}_{U+L}^{\gamma^*g } -y^2 \,{\cal A}_{L}^{\gamma^*g}} \,\frac{h_1^{\perp\, g}(x,\bm q_\sT^2 )}{ f_1^{g}(x,\bm q_\sT^2 )}\,.
\label{eq:cos2phiT2phiP}
\end{align}

The gluon TMDs for a transversely polarized proton can be extracted in a similar fashion. In particular, the polarized cross section, Eq.~(\ref{eq:csT}), can be simplified by integrating over the angle $\phi_\perp$, yielding:
\begin{align}
\int\d\phi_\perp\d\sigma^T & =  2\pi {{\cal N}\,\vert \bm S_\sT\vert}\, \frac{ |\bm q_\sT | }{M_p} \left [\mathcal{A}^{eg}_0 \sin(\phi_S -\phi_\sT)    f_{1T}^{\perp\, g} (x,\bm{q}_{\sT}^2) -\frac{1}{2} \mathcal{B}^{eg}_0\sin(\phi_S-3\phi_\sT)   \frac{ |\bm q_\sT |^2 }{M_p^2} \,h_{1T}^{\perp\, g} (x,\bm{q}_{\sT}^2) \right . \nonumber\\*
  &\qquad+\mathcal{B}_0^{eg} \sin(\phi_S+\phi_\sT)  h_{1}^{g} (x,\bm{q}_{\sT}^2) \bigg ]\,,
\label{eq:csTh}
\end{align}
where we have introduced the combination $h_1^g$ of Eq.\ (\ref{eq:h1}). Clearly, the resulting integrated cross section has only three independent azimuthal modulations left, each related to a different T-odd gluon TMD. Note that this situation is analogous to the case of quark azimuthal asymmetries in SIDIS ($e\, p^\uparrow \to e' \, h\, X$), in which the role of $\phi_T$ is played by $\phi_h$~\cite{Boer:1997nt}.

Using Eq.~(\ref{eq:A}), one then obtains (setting $\vert \bm S_\sT\vert=1$)
\begin{align}
A^{\sin(\phi_S-\phi_\sT)} & =  \frac{\vert \bm q_\sT\vert}{M_p}\, \frac{ f_{1T}^{\perp\,g}(x,\bm q_\sT^2) }{f_1^g(x,\bm q_\sT^2)}\,,
\label{eq:A1}\\
A^{\sin(\phi_S+\phi_\sT)}  & =  \frac{\vert \bm q_\sT\vert}{M_p}\,\frac{{\cal B}_0^{eg}}{{\cal A}_0^{eg}} \frac{ h_{1 }^{g}(x,\bm q_\sT^2) }{f_1^g(x,\bm q_\sT^2)}\,,
 \label{eq:A2}\\
A^{\sin(\phi_S-3\phi_\sT)}  & =   -  \frac{\vert \bm q_\sT\vert ^3}{2M_p^3}\, \frac{{\cal B}_0^{eg}}{{\cal A}_0^{eg}} \frac{ h_{1T}^{\perp\,g}(x,\bm q_\sT^2)}{f_1^g(x,\bm q_\sT^2)} \label{eq:A3}\,,
\end{align}
where we note that the asymmetries in Eqs.~(\ref{eq:A2}) and (\ref{eq:A3}) vanish when $y\to 1$, see the last line of Eq.~\eqref{eq:Agstar},  corresponding to a longitudinally polarized virtual photon. Furthermore, as already pointed out in Ref.~\cite{Bacchetta:2018ivt} for the process $e\, p \to e\, J/\psi\, X$, the following ratios of asymmetries
\begin{align}
\frac{A^{\cos2\phi_\sT }}{A^{\sin(\phi_S+\phi_\sT)}}= \frac{\vert\bm q_\sT\vert}{M_p}\, \frac{h_{1}^{\perp\,g}(x,\bm q_\sT^2)}{h_{1}^{g}(x,\bm q_\sT^2)}\,,
\label{eq:ratioA1}\\
\frac{A^{\sin(\phi_S-3\phi_\sT)}}{A^{\cos 2\phi_\sT }}=- \frac{\vert \bm q_\sT\vert}{2 M_p}\, \frac{h_{1 \sT}^{\perp\,g}(x,\bm q_\sT^2)}{h_{1}^{\perp\,g}(x,\bm q_\sT^2)}\,,\\
\frac{A^{\sin(\phi_S-3\phi_\sT)}}{A^{\sin(\phi_S+\phi_\sT)}} =- \frac{\bm q_\sT^2}{2 M_p^2}\, \frac{h_{1T}^{\perp\,g}(x,\bm q_\sT^2)}{h_{1}^{g}(x,\bm q_\sT^2)}\, \,.
\label{eq:ratioA3}
\end{align}
are directly sensitive to the relative magnitude of the various gluon TMDs, without any dependence either on the different LDMEs or on the other kinematic variables in the process.

\subsection{Use of positivity bounds}
In order to assess the maximum allowed size of the above asymmetries, one can make use of the fact that the polarized gluon TMDs have to satisfy the following, model independent, positivity bounds~\cite{Mulders:2000sh}:
\begin{align}
\frac{\vert \bm q_\sT \vert }{M_p}\, \vert f_{1T}^{\perp \,g}(x,\bm q_\sT^2) \vert & \le   f_1^g(x,\bm q_\sT^2)\,,\nonumber \\
\frac{ \bm q^2_\sT }{2 M_p^2}\, \vert h_{1}^{\perp\,g}(x,\bm q_\sT^2) \vert & \le   f_1^g(x,\bm q_\sT^2)\,,\nonumber \\
\frac{\vert \bm q_\sT \vert }{M_p}\, \vert h_{1}^g(x,\bm q_\sT^2) \vert & \le   f_1^g(x,\bm q_\sT^2)\,,\nonumber \\
\frac{\vert \bm q_\sT \vert^3}{2 M_p^3}\, \vert h_{1T}^{\perp \,g}(x,\bm q_\sT^2) \vert & \le   f_1^g(x,\bm q_\sT^2)\,.
\label{eq:bounds}
\end{align}
Applying these to the asymmetries in Eqs.~(\ref{eq:cos2phiT}) and (\ref{eq:cos2phiT2phiP}), we find that:
\begin{equation}
\left \vert A^{\cos 2\phi_{\sT}} \right \vert \leq\frac{2{\vert \cal B}_0^{e g}\vert }{{\cal A}_0^{e g}}\qquad \mathrm{and}\qquad
\left \vert A^{\cos 2(\phi_{\sT}-\phi_\perp)} \right \vert \leq \frac{2\vert {\cal B}_2^{e g} \vert}{{\cal A}_0^{e g}}\;,
\label{eq:assybounds}
\end{equation}
where the maxima on the r.h.s. are always independent of the center-of-mass energy. As can be seen from Eqs. (\ref{eq:A2}) and (\ref{eq:A3}), the upper bounds of the asymmetries $A^{\sin(\phi_S+\phi_\sT)}$ and $A^{\sin(\phi_S-3\phi_\sT)}$ are equal to half the one for $A^{\cos 2 \phi_{\sT}}$, while the bound for the Sivers asymmetry $A^{\sin(\phi_S-\phi_\sT)}$, Eq.~(\ref{eq:A1}), is simply equal to one.

\subsection{McLerran-Venugopalan model}
As an alternative to the model-independent positivity bounds in the previous section, the gluon TMDs can be evaluated in a model, in order to show predictions for the azimuthal asymmetries in Eqs.~(\ref{eq:cos2phiT}) and (\ref{eq:cos2phiT2phiP}). From the expression of $x$, which can be read off the delta function in Eq.~(\ref{eq:delta}), it follows that for realistic EIC center-of-mass energies $\sqrt{s}\sim\,100\,\mathrm{GeV}$, and for values of $y$ that are not too small, its value is of the order of $x\sim 10^{-2}$. In this regime, the nonperturbative McLerran-Venugopalan (MV) model \cite{McLerran:1993ni,McLerran:1993ka,McLerran:1994vd} for the gluon distributions inside an unpolarized nucleus  is phenomenologically very successful. In view of the total lack of information on gluon TMDs, following Ref.~\cite{Boer:2016fqd}, we will apply it also to the proton case. The unpolarized and linearly polarized gluon TMDs in the MV model read \cite{Metz:2011wb,Dominguez:2011br}:
\begin{align}
f_1^g(x,\bm{q}_\sT^2)=\frac{S_\perp C_F}{\alpha_s \pi^3} \int \mathrm{d}r \frac{J_0(q_\sT r)}{r}\left (1-e^{-\frac{r^2}{4}Q_{sg}^2(r) }\right )\,,\\
h_1^{\perp g}(x,\bm{q}_\sT^2)=\frac{S_\perp C_F}{\alpha_s \pi^3} \frac{2 M_p^2}{\bm{q}_\sT^2} \int \mathrm{d}r \frac{J_2(q_\sT r)}{r \ln \frac{1}{r^2 \Lambda^2}}\left (1-e^{-\frac{r^2}{4}Q_{sg}^2(r) }\right )\,.
\end{align}
In the above formulas, $S_\perp$ is the transverse size of the proton and $\Lambda$ is an infrared cutoff (we take $\Lambda = \Lambda_\mathrm{QCD}$ = 0.2~GeV). Furthermore, $Q_{sg}(r)$ is the gluon saturation scale, which we parameterize, as is usually done in the MV model, as $Q\bm_{sg}^2(r)=Q_{sg0}^2 \ln{(1/r^2 \Lambda^2)} $. Finally, the gluon saturation scale is related to the one felt by the quarks by a color factor, $Q_{sg0}^2=(N_c/C_F) Q_{sq0}^2$, and we take the numerical value $Q_{sq0}^2=0.35\,\mathrm{GeV}^2$ at $x=10^{-2}$ from the fit of Ref.~\cite{GolecBiernat:1998js}.

The ratio of $h^{\perp g}_1$ over $f^g_1$ is therefore:
\begin{align}
\frac{\bm{q}_{\sT}^{2}}{2M_{p}^{2}}\frac{h_{1}^{\perp g}(x,\bm{q}_{\sT}^{2})}{f_{1}^{g}(x,\bm{q}_{\sT}^{2})}=
\frac{\int\mathrm{d}r\frac{J_{2}(q_{\sT}r)}{r\ln(\frac{1}{r^{2}\Lambda^{2}}+e)}\left (1-e^{-\frac{r^{2}}{4}Q_{sg0}^{2}\ln \left (\frac{1}{r^{2}\Lambda^{2}}+e \right )}\right )}{\int\mathrm{d}r\frac{J_{0}(q_{\sT}r)}{r}\left (1-e^{-\frac{r^{2}}{4}Q_{sg0}^{2}\ln \left (\frac{1}{r^{2}\Lambda^{2}}+e \right )}\right )}\,,
\label{eq:ratioanalytic}
\end{align}
where $e$ was added as a regulator to guarantee numerical convergence.
In the following section, the above model is used to show the $q_\sT$ dependence of the asymmetries in Eqs. (\ref{eq:cos2phiT}) and (\ref{eq:cos2phiT2phiP}).

\begin{table}[t]
\begin{centering}
\begin{tabular}{|c|c|c|c|c|c|}
\hline
$J/\psi$ & $\langle0|\mathcal{O}_{8}^{J/\psi}\bigl(^{1}S_{0}\bigr)|0\rangle$ & $\langle0|\mathcal{O}_{8}^{J/\psi}\bigl(^{3}S_{1}\bigr)|0\rangle$& $\langle0|\mathcal{O}_{1}^{J/\psi}\bigl(^{3}S_{1}\bigr)|0\rangle$ & $\langle0|\mathcal{O}_{8}^{J/\psi}\bigl(^{3}P_{0}\bigr)|0\rangle/m_c^2$ & \tabularnewline
\hline
\hline
Sharma et al. \cite{Sharma:2012dy}& $1.8 \pm 0.87 $ & $0.13 \pm 0.13 $& $1.2 \times10^{2} $& $1.8 \pm 0.87$ & $\times10^{-2}\,\mathrm{GeV}^{3}$\tabularnewline
Chao et al. \cite{Chao:2012iv} &$8.9 \pm 0.98 $ & $0.30 \pm 0.12 $& $1.2 \times10^{2} $& $0.56 \pm 0.21$ & $\times10^{-2}\,\mathrm{GeV}^{3}$\tabularnewline
\hline
\end{tabular}
\par\end{centering}
\caption{Numerical values of the LDMEs for $J/\psi$ production.}
\label{tab:jpsiLDME}
\end{table}
\begin{table}[t]
\begin{centering}
\begin{tabular}{|c|c|c|c|c|c|}
\hline
$\Upsilon(1S)$ & $\langle0|\mathcal{O}_{8}^{\Upsilon}\bigl(^{1}S_{0}\bigr)|0\rangle$ & $\langle0|\mathcal{O}_{8}^{\Upsilon}\bigl(^{3}S_{1}\bigr)|0\rangle$& $\langle0|\mathcal{O}_{1}^{\Upsilon}\bigl(^{3}S_{1}\bigr)|0\rangle$ & $\langle0|\mathcal{O}_{8}^{\Upsilon}\bigl(^{3}P_{0}\bigr)|0\rangle/5m_b^2$ & \tabularnewline
\hline
\hline
Sharma et al. \cite{Sharma:2012dy}& $1.2 \pm 4.0 $ & $4.8 \pm 3.3 $& $11 \times10^{2} $& $1.2 \pm4.0$ & $\times10^{-2}\,\mathrm{GeV}^{3}$\tabularnewline
\hline
\end{tabular}
\par\end{centering}
\caption{Numerical values of the LDMEs for $\Upsilon$ production.}
\label{tab:upsilonLDME}
\end{table}

\section{Numerical results}
\label{results}
Before presenting our results, we come back to the assumption of negligible quark contribution. To this aim, we compute the ratio of the collinear quark- over total unpolarized cross section in NRQCD, using for the parton distribution functions (PDFs) the MSTW2008LO set~\cite{Martin:2009iq} and adopting the same LDME sets as for the study of the azimuthal asymmetries, see Table \ref{tab:jpsiLDME} (with the charm mass $m_c=1.3\,\mathrm{GeV}$)
for $J/\psi$ and Table \ref{tab:upsilonLDME} (with the bottom mass $m_b=4.2\,\mathrm{GeV}$)
for $\Upsilon$. Following Ref.~\cite{Kniehl:2001tk}, we choose the hard scale for these PDFs to be equal to $\xi \sqrt{M^2+Q^2}$, where $\xi$ varies between $1/2$ and $2$, and take a conservative estimate for the center-of-mass energy at the EIC, i.e.~$\sqrt{s}=65\;\mathrm{GeV}$. As can be clearly seen in every plot we present in what follows, in the kinematic regions where the asymmetries can be sizable, the estimated quark contribution to the unpolarized cross section is always practically negligible, at least at our current level of accuracy.

\begin{figure}[t]
\begin{center}
\includegraphics[scale=.8]{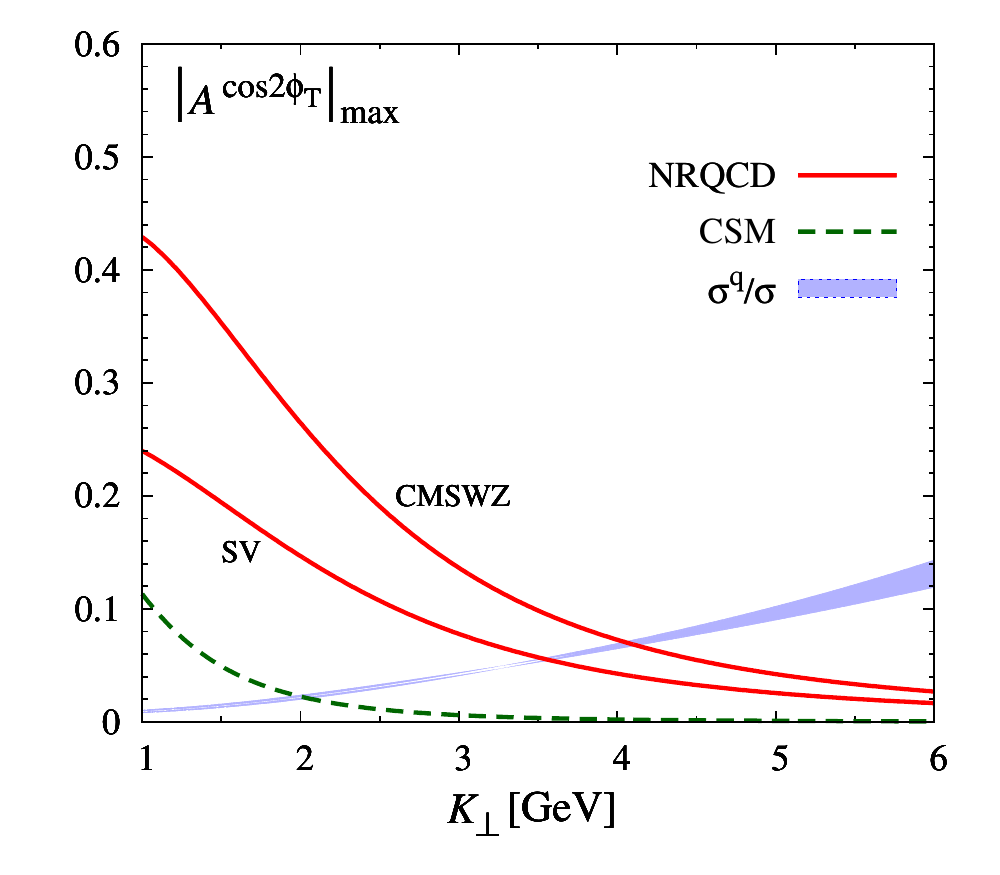}
\includegraphics[scale=.8]{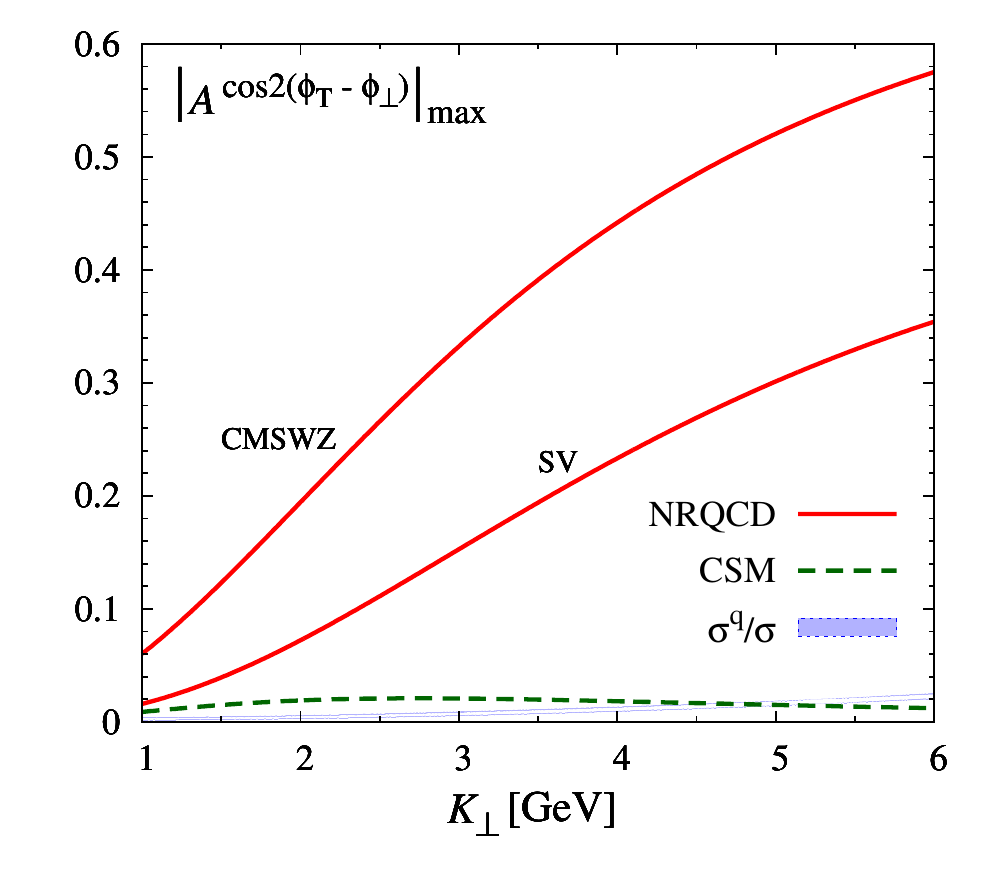}
\end{center}
\caption{Upper bounds for the absolute values of $A^{\mathrm{cos} 2 \phi_\sT}$ (left) and $A^{\mathrm{cos}\,2 (\phi_T-\phi_\perp)}$ (right) for the $J/\psi$, in NRQCD and in the CSM, as a function of $K_\perp$ at $Q^2=10\,\mathrm{GeV}^2$. We take $z=0.7$, $y=0.3$ (left) or $z=0.3$, $y=0.7$ (right). Two sets of LDMEs are used: SV \cite{Sharma:2012dy} and CMSWZ \cite{Chao:2012iv}. The ratio of the collinear quark contribution over the total unpolarized cross section is shown in the form of a band, representing the scale uncertainty and the results obtained with the two LDME sets, for the center-of-mass energy $\sqrt{s}=65\; \mathrm{GeV}$.}
\label{fig:bound_Kt}
\end{figure}

\begin{figure}[b]
\begin{center}
\includegraphics[scale=.8]{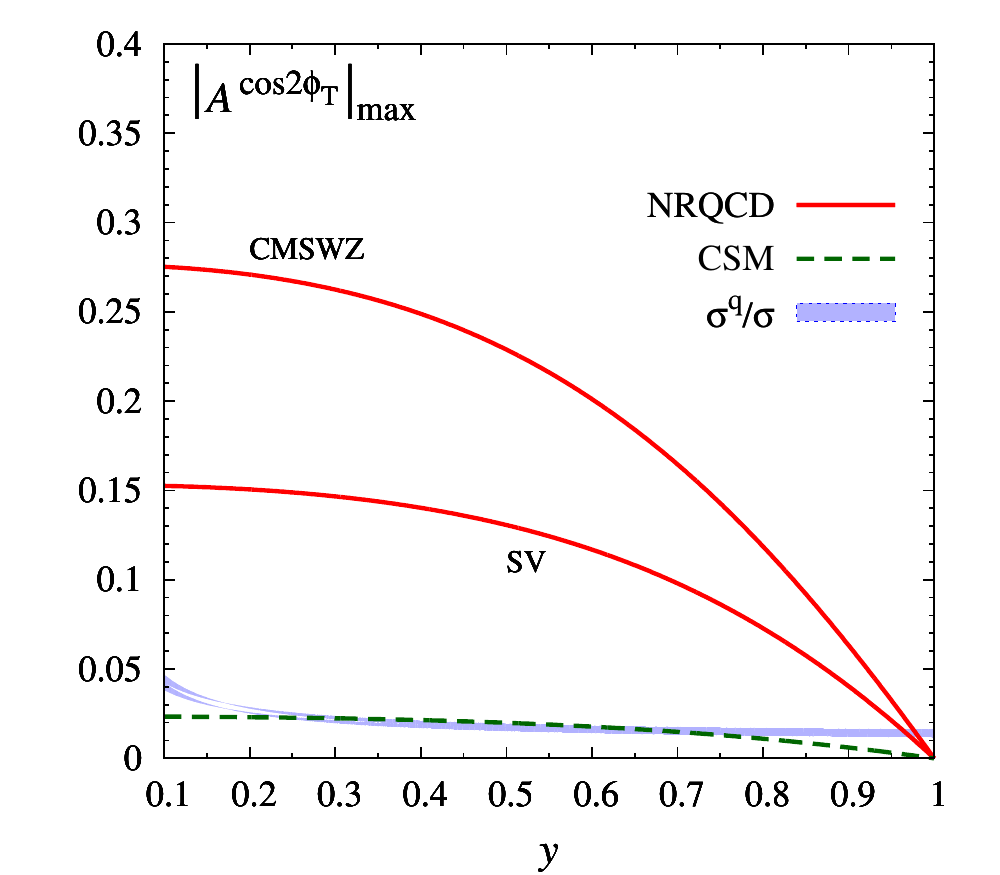}
\includegraphics[scale=.8]{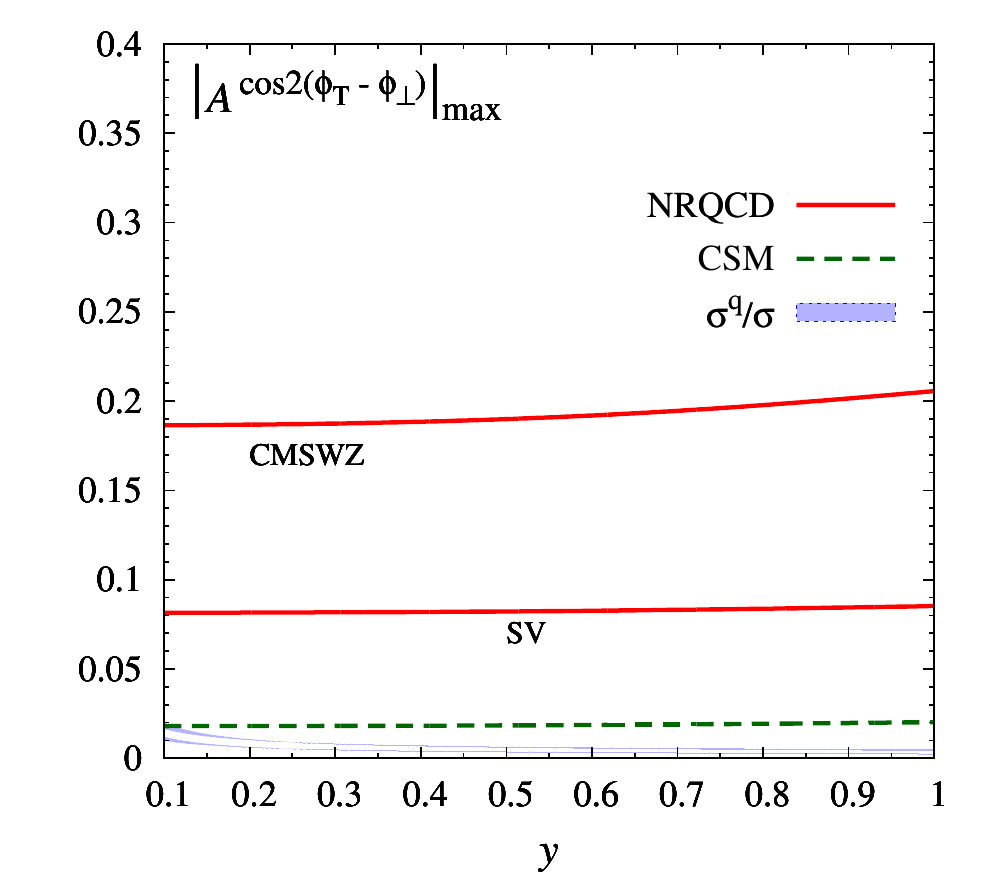}
\end{center}
\caption{Upper bounds for the absolute values of $A^{\mathrm{cos} 2 \phi_\sT}$ (left) and $A^{\mathrm{cos}\,2 (\phi_T-\phi_\perp)}$ (right) for the $J/\psi$, in NRQCD and in the CSM, as a function of $y$ at $Q^2=10\,\mathrm{GeV}^2$ and $K_\perp=2\,\mathrm{GeV}$. We take $z=0.7$ (left) or $z=0.3$ (right). Two sets of LDMEs are used: SV \cite{Sharma:2012dy} and CMSWZ \cite{Chao:2012iv}. The ratio of the collinear quark contribution over the total unpolarized cross section is shown in the form of a band, representing the scale uncertainty and the results obtained with the two LDME sets, for the center-of-mass energy $\sqrt{s}=65\; \mathrm{GeV}$.
}
\label{fig:bound_y}
\end{figure}

To start, in Fig.~\ref{fig:bound_Kt} we show the upper bounds of the absolute values of the asymmetries $A^{\mathrm{cos} 2 \phi_\sT}$ (left panel) and $A^{\mathrm{cos} 2 (\phi_\sT-\phi_\perp)}$ (right panel), for the  $J/\psi$ meson, as a function of the transverse momentum $K_\perp$. We identified two regions in $z$ and $y$ where either of them is very large, corresponding to, respectively, $z=0.7$ and $y=0.3$, or $z=0.3$ and $y=0.7$.  The $J/\psi$ mass is taken to be $M=3.1\;\mathrm{GeV}$, and the photon virtuality is fixed at $Q^2=10\,\mathrm{GeV}^2$ in order to keep all the large scales, i.e. $K_\perp$, $M$ and $Q$, in the same ballpark. The full lines  correspond to the complete NRQCD calculation, while the CS results are shown as dashed lines. For these asymmetries, NRQCD consistently leads to larger upper bounds as compared to the CSM, and in certain kinematic regions even to a drastically different behavior (see right panel). Indeed, in contrast to the CSM, the Color Octet mechanism generates a large upper bound for the $A^{\mathrm{cos} 2 (\phi_\sT-\phi_\perp)}$ asymmetry. Furthermore, the NRQCD results exhibit a strong dependence on the choice of the different LDME sets.  Clearly, our process can lead to large asymmetries, with upper bounds in certain regions reaching almost $60\%$. Notice that we have selected specific kinematic ranges where the asymmetries could be potentially very large and where, consequently, any measurement could unambiguously allow to put strong constraints on the TMDs under study.

\begin{figure}[t]
\begin{center}\hspace*{-8.5cm}
\includegraphics[scale=.8]{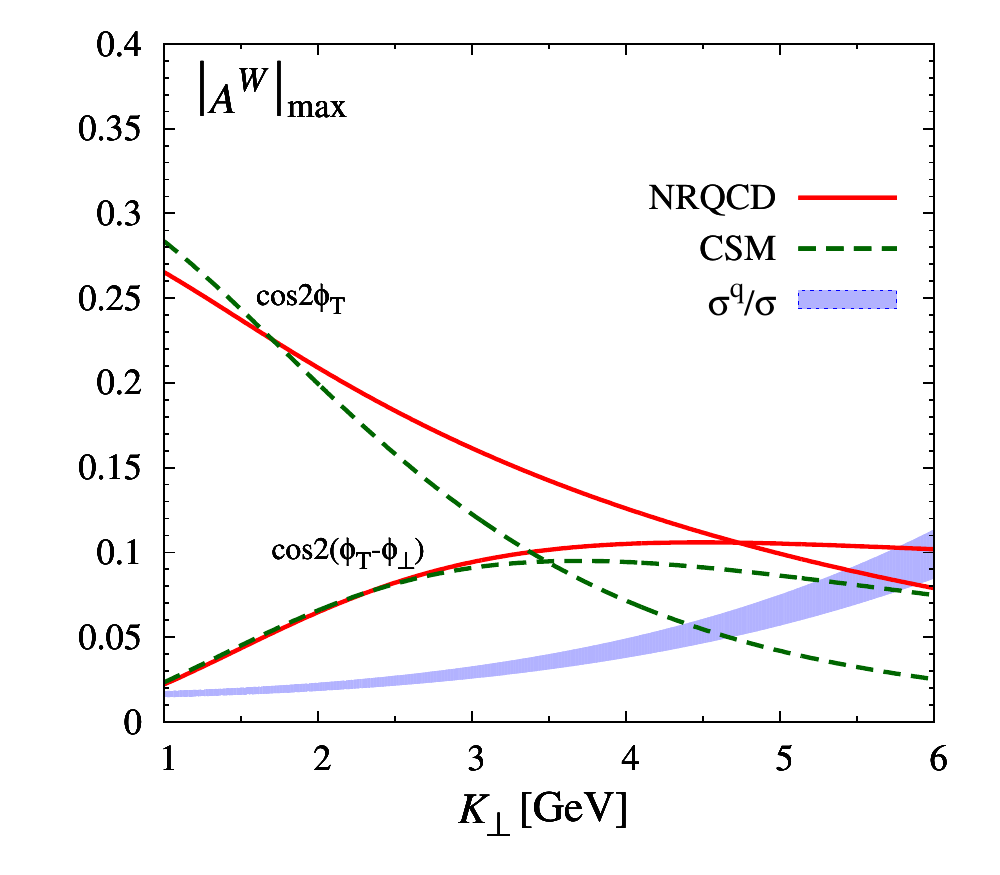}\\
\vspace{-7.1cm}\hspace*{8.5cm}
\includegraphics[scale=.755]{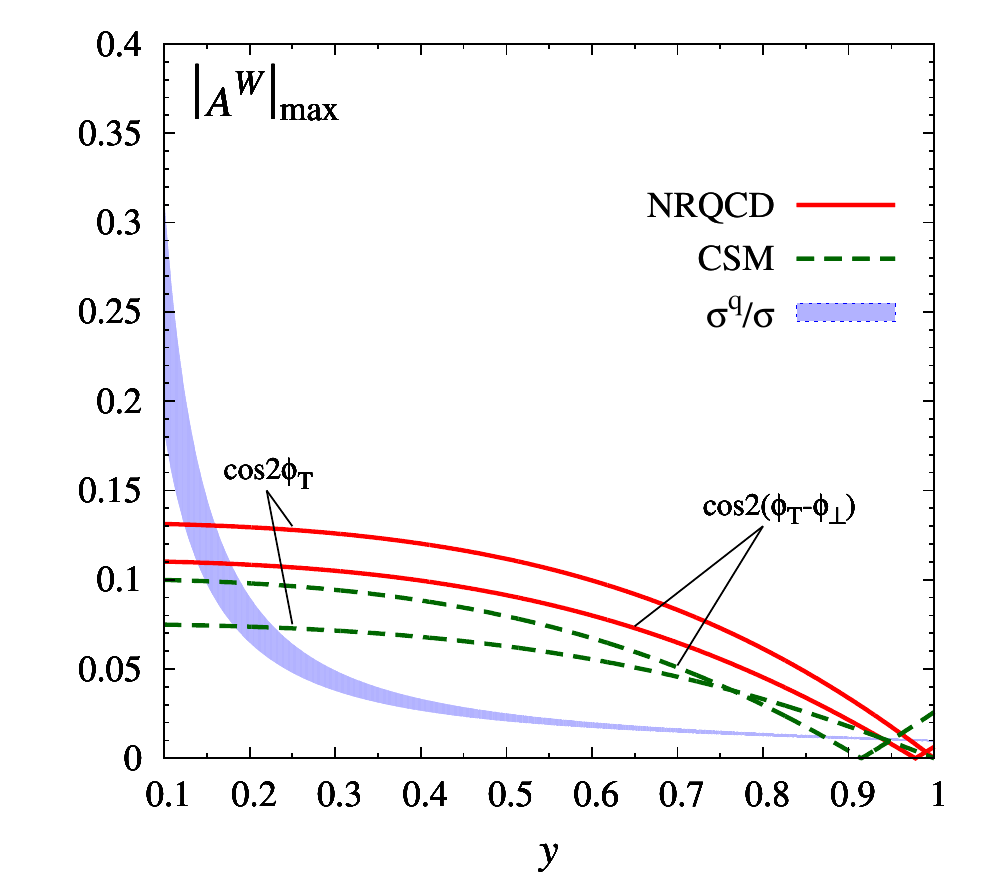}
\end{center}
\caption{Upper bounds for the absolute values of $A^W$, with $W=\mathrm{cos}\,2 \phi_\sT,\,\mathrm{cos}\,2 (\phi_T-\phi_\perp)$ for the $\Upsilon$, in NRQCD and in the CSM, at $Q^2=100\,\mathrm{GeV}^2$, as a function of $K_\perp$ at $y =0.3$ and $z=0.7$ (left) and as a function of $y$ at $K_\perp = 4$ GeV and $z=0.6$ (right). The SV set of LDMEs~\cite{Sharma:2012dy} is used. The ratio of the collinear quark contribution over the total unpolarized cross section is shown in the form of a band, representing the scale uncertainty, for the center-of-mass energy $\sqrt{s}=65\; \mathrm{GeV}$.}
\label{fig:Ups_bound_Kt}
\end{figure}

\begin{figure}[b]
\begin{center}
\includegraphics[scale=.8]{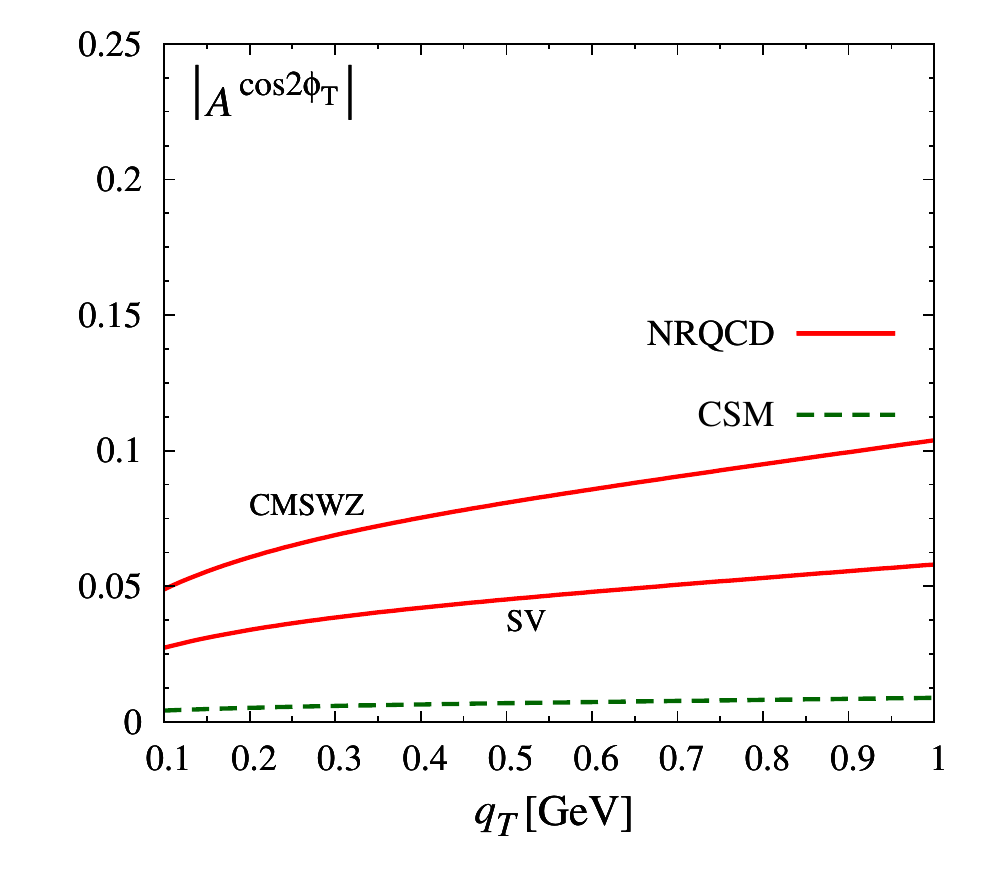}
\includegraphics[scale=.8]{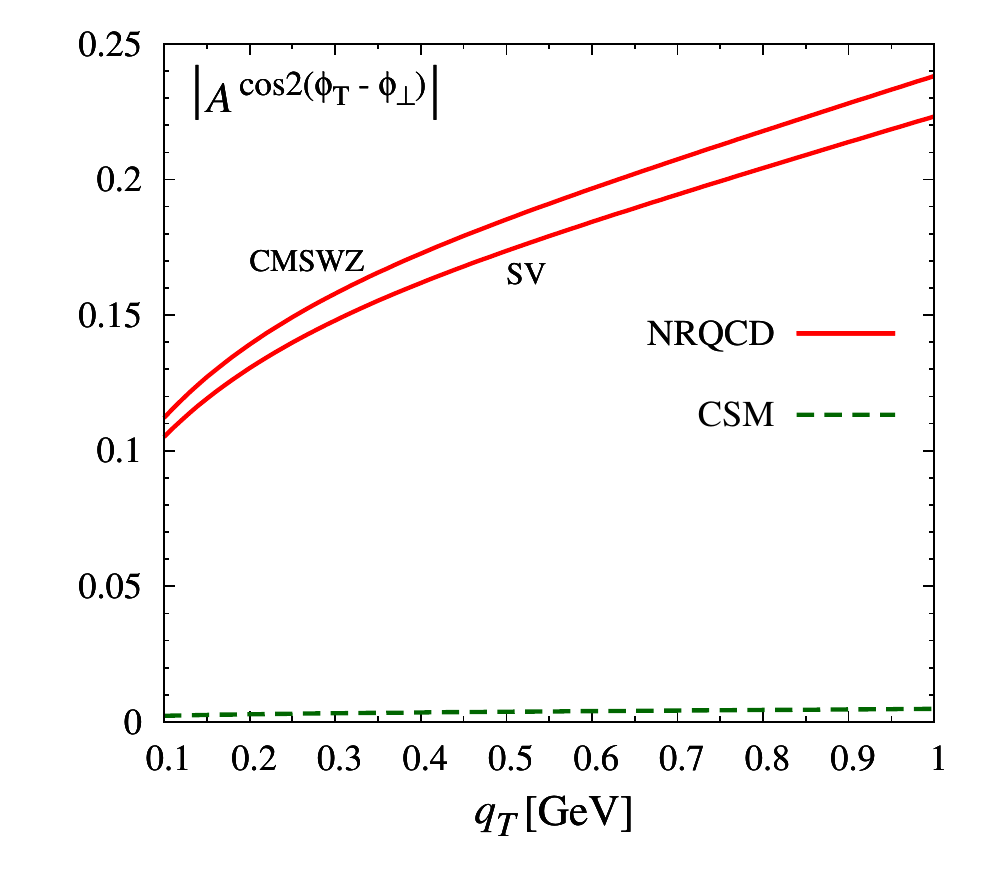}
\end{center}
\caption{Absolute values of $A^{\cos 2 \phi_\sT}$ (left) and $A^{\cos 2 (\phi_T-\phi_\perp)}$ (right) for $J/\psi$ production as a function of $q_\sT$, in the MV model, for $Q^2={10}\,\mathrm{GeV}^2$, $z=0.7$, $y=0.3$, $K_\perp=2\,\mathrm{GeV}$ (left) and  $z=0.3$, $y=0.7$, $K_\perp=6\,\mathrm{GeV}$ (right).}
\label{fig:MV}
\end{figure}

The $y$ dependence of the above asymmetries for $J/\psi$ production is presented in Fig.~\ref{fig:bound_y}, for a fixed value of $K_\perp=2\,\mathrm{GeV}$, and again both for the full NRQCD calculation (solid lines) and the CSM (dashed lines). Since $\sqrt{s}$ is fixed,  $y$ is consistently cut at $y =0.1$ to fulfill all kinematic constraints. In the opposite limit,  $y\to 1$, the $A^{\cos{2\phi_\sT}}$ asymmetries vanish, as expected from the expression in Eq.~(\ref{eq:cos2phiT}). The maximum of $A^{\cos{2(\phi_\sT-\phi_\perp)}}$, on the other hand, is almost independent of $y$.

In Fig.~\ref{fig:Ups_bound_Kt}, the upper bounds for the same asymmetries are presented for the $\Upsilon(1S)$ quarkonium state, whose mass is taken to be $M=9.5\,\mathrm{GeV}$, and for which the single LDME set available, see Table \ref{tab:upsilonLDME}, is used (taking $m_b = 4.2$ GeV). Moreover, to avoid potentially large logarithmic effects, we choose a virtuality of the same order as the $\Upsilon$ mass: $Q^2=100\,\mathrm{GeV}^2$. In contrast to the $J/\psi$ case, we find that the bounds for the asymmetries are only sizable in the region of intermediate and low values of $y$ and large values of $z$. Notice that below $y=0.15$ (right panel) the quark contribution to the unpolarized cross section could become important and no more negligible.

As is evident from Eqs.~(\ref{eq:cos2phiT}) and (\ref{eq:cos2phiT2phiP}), the asymmetries $A^{\cos 2 \phi_\sT}$ and $A^{\cos 2(\phi_\sT-\phi_\perp)}$ are sensitive to the ratio of linearly polarized and unpolarized gluon TMDs. As discussed in the previous section, in the case of a large nucleus (and at low $x$), analytical expressions for these TMDs are available within the MV model. Adopting this model also for a proton target and at $x\simeq 10^{-2}$, allows us to calculate the $q_\sT$ dependence of the above asymmetries, for fixed values of the other variables. As can be seen in Fig.~\ref{fig:MV} for $J/\psi$ production, using $Q^2=10\,\mathrm{GeV}^2$ everywhere, and $K_\perp=2\,\mathrm{GeV}$, $z=0.7$, and $y=0.3$ (left panel) or $K_\perp=6\,\mathrm{GeV}$, $z=0.3$, and $y=0.7$ (right panel), the corresponding asymmetries in the full NRQCD calculation can still be large, particularly $A^{\cos 2(\phi_\sT-\phi_\perp)}$.

\begin{figure}[t]
\begin{center}
\includegraphics[scale=.8]{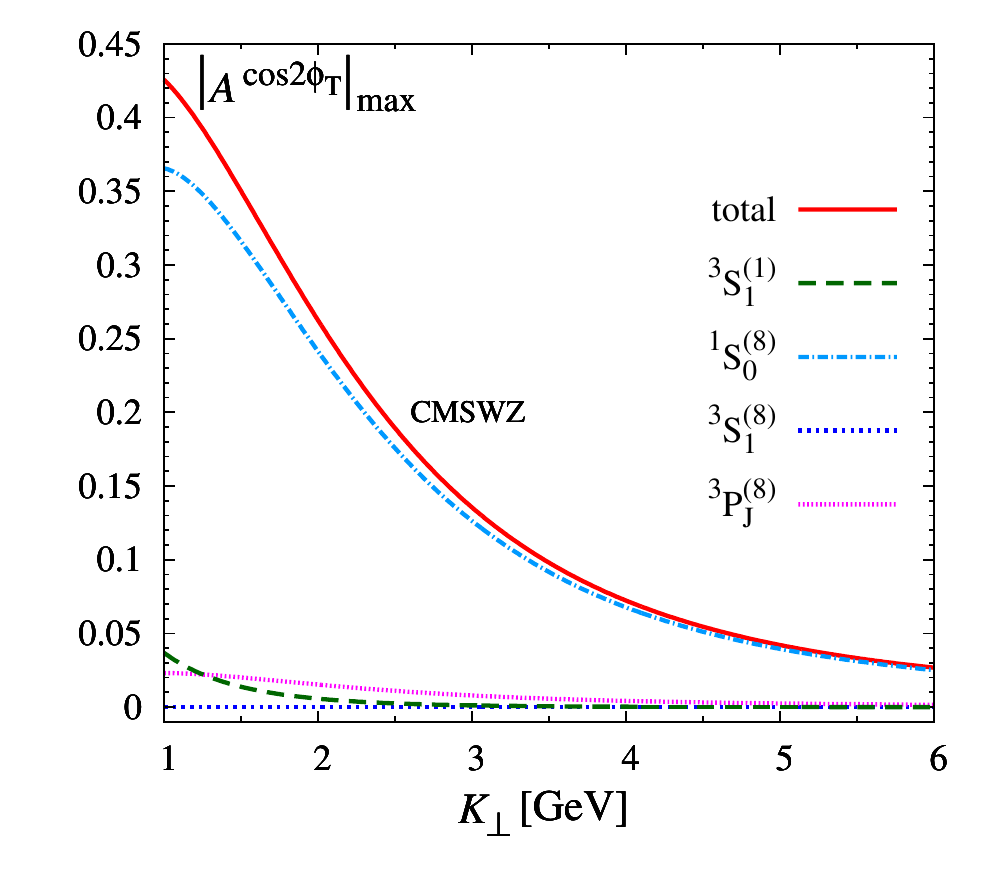}
\includegraphics[scale=.8]{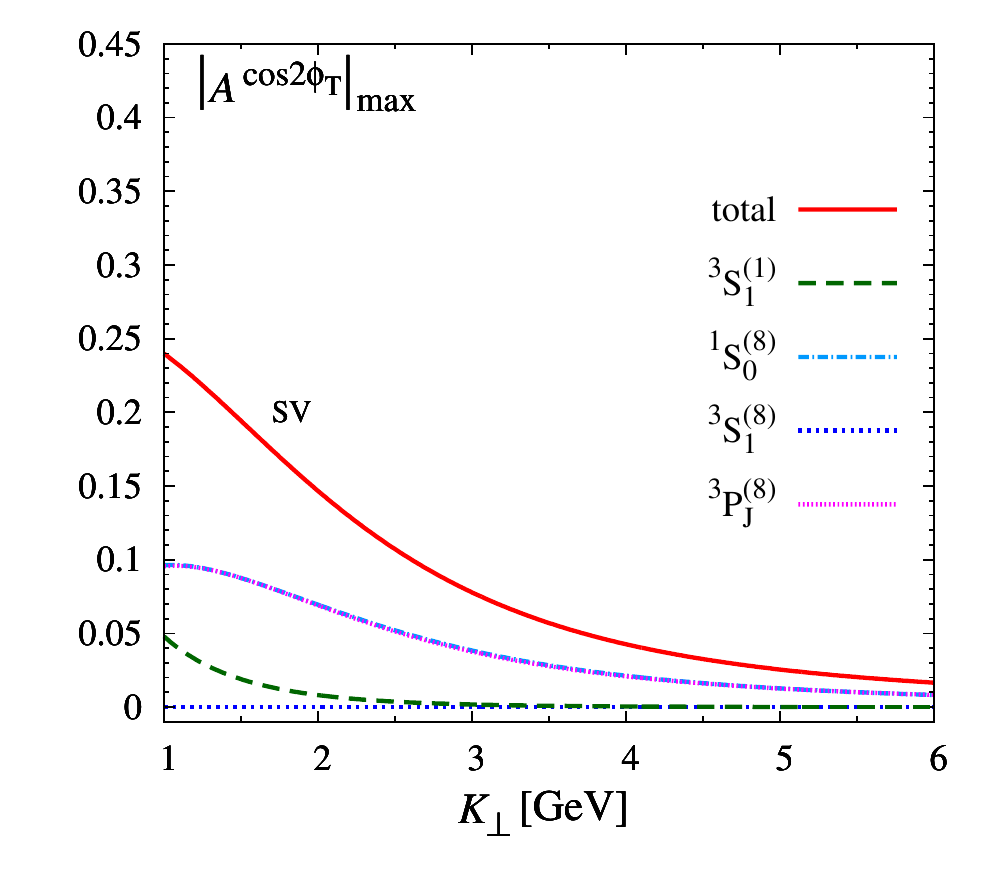}
\end{center}
\caption{Upper bound for the absolute values of $A^{\cos 2 \phi_\sT}$ for $J/\psi$, at $Q^2={10}\,\mathrm{GeV}^2$, $z=0.7$, $y=0.3$, for the LDME sets CMSWZ (left) and SV (right). The upper bound is decomposed in all the contributions to the total NRQCD result, i.e.~the $^1S^{(8)}_0$, $^3S^{(8)}_1$, and $^3P^{(8)}_J$ waves in the Color Octet, and the $^3S^{(1)}_1$ wave in the Color Singlet.}
\label{fig:decomposition}
\end{figure}

Finally, in Fig.~\ref{fig:decomposition} we illustrate the NRQCD decomposition in its different waves for the upper bound of $A^{\cos{2\phi_\sT}}$ for $J/\psi$. As one can see, for the CMSWZ set (left panel) there is a clear dominance of one single wave: $^1S^{(8)}_0$, while for the SV set there are two dominant waves which contribute almost equally, i.e.\ $^1S^{(8)}_0$ and $^3P^{(8)}_J$. This behavior could be traced back to the values of the corresponding LDMEs.

\section{Conclusions and outlook}
\label{conclusions}
In this work, we have studied the quarkonium + jet electroproduction under the assumption of TMD factorization, within NRQCD. We have been able to identify broad kinematic regions where the quark contribution is negligible, and where the cross section can be analyzed only in terms of gluon TMDs.
The specific azimuthal modulations entering the cross sections could allow for a direct access to important TMDs that are still completely unknown, like those involving linearly polarized gluons in unpolarized or transversely polarized nucleons.

With the help of positivity bounds for these TMDs we have demonstrated that, over a range of accessible regions of the phase space, the arising azimuthal asymmetries are potentially very large. Consequently, this process could be an excellent way to experimentally access the ratio of the linearly polarized gluon TMD over the unpolarized one at a future Electron-Ion Collider.

Besides these model independent estimates, we have also considered a nonperturbative model, valid in the small-$x$ regime, namely the MV model, and presented some predictions which could be tested against experimental data.

Like any leading-order calculation, our work can be improved upon in different ways. Notably, should one aim to reliably extract gluon TMDs from precision data (at present not available) on the process under study, our computation of the cross section could be extended to take the following aspects into account. Firstly, the hadronization of the heavy-quark pair state into the quarkonium involves the emission of soft gluons, which can smear its transverse momentum over a range of the order of $2 m_c v$, an effect which can be encoded in the so-called quarkonium TMD shape functions recently introduced in~\cite{Echevarria:2019ynx,Fleming:2019pzj}. Likewise, the precision with which the transverse momentum of the gluon can be reconstructed is unavoidably limited by a factor of order $\Lambda_{\mathrm{QCD}}$ due to hadronization. Moreover, an additional uncertainty can be expected due to wide-angle radiation that escapes the jet algorithm, and hence causes a mismatch between the momentum of the reconstructed jet and the initiating gluon. As was pointed out in Refs.~\cite{Buffing:2018ggv, Gutierrez-Reyes:2018qez, Gutierrez-Reyes:2019msa}, a part of this mismatch can be accounted for with the inclusion of TMD jet functions in the cross section.

In order to be able to include the above mentioned improvements to our computation in a systematic way, in addition to the Sudakov resummations and the higher order contributions in perturbation theory, it would be highly desirable to have a proof of TMD factorization for this process (e.g. in analogy with the one recently formulated for $p\,p\to \eta_cX$~\cite{Echevarria:2019ynx}). Such a factorization theorem could also shed light on the precise role of the CO long-distance matrix elements at small transverse momenta, and pave the way to a dedicated extraction of them in the TMD regime. Indeed, at our present precision, the largest source of theoretical uncertainty comes from the LDMEs which, although they are expected to be universal, vary significantly among different (collinear) fits.

We believe that this analysis represents an important step towards a better understanding of gluon TMDs. Moreover, our findings could be relevant for the study of their process dependence, in particular when compared to similar analyses for proton-proton collisions, which could be performed at various ongoing or planned experiments at RHIC and LHC~\cite{Brodsky:2012vg,Hadjidakis:2018ifr,Aidala:2019pit}.

\acknowledgments
U.D.~acknowledges partial support by Fondazione Sardegna under the project {\em Quarkonium at LHC energies}, CUP
F71I17000160002 (University of Cagliari).

\end{document}